\documentclass[fleqn,usenatbib]{mnras}

\usepackage{newtxtext,newtxmath,dsfont}
\usepackage[T1]{fontenc}

\DeclareRobustCommand{\VAN}[3]{#2}
\let\VANthebibliography\thebibliography
\def\thebibliography{\DeclareRobustCommand{\VAN}[3]{##3}\VANthebibliography}

\usepackage{graphicx}	% Including figure files
\usepackage{amsmath}	% Advanced maths commands
\usepackage{tikz}
\usetikzlibrary{shapes.geometric, arrows.meta}
\tikzstyle{io} = [trapezium, trapezium left angle=70, trapezium right angle=110, minimum width=1.5cm, minimum height=1cm, text centered, draw=black, fill=blue!30, text width=2cm]
\tikzstyle{process} = [rectangle, minimum width=1.5cm, minimum height=1cm, text centered, draw=black, fill=orange!30, text width=2.5cm]
\tikzstyle{decision} = [diamond, minimum width=1cm, minimum height=.5cm, text centered, draw=black, fill=green!30, text width=2cm]
\tikzstyle{vertex} = [text centered] % sleeker? 
\tikzstyle{arrow} = [thick,->,>=stealth]

\usepackage{color} % to allow new color definitions
\definecolor{bleudefrance}{rgb}{0.19, 0.55, 0.91}
\definecolor{AmericanRed}{rgb}{0.698, 0.133, 0.204}
\definecolor{AmericanBlue}{rgb}{0.0391, 0.1914, 0.3789}
\definecolor{mustard}{rgb}{.808,.702,.00392}
\definecolor{ketchup}{rgb}{0.781, 0.160, 0.1328}

\newcommand*{\wm}[1]{\textcolor{AmericanRed}{\textsf{#1}}}

\newcommand*{\gus}[1]{\textcolor{bleudefrance}{\textsf{#1}}}

\newcommand{\SpARCS}{The \textit{Spitzer} Adaptation of the Red-sequence Cluster Survey} % SpARCS \citep[\SpARCS][]{Wilson+09,Muzzin+09}

\DeclareMathOperator{\erf}{erf} % error function 

\title[Red Dragon]{Red Dragon: A Redshift-Evolving Gaussian Mixture Model for Galaxies}
\author[W. K. Black et al.]{
\href{https://orcid.org/0000-0003-4811-7913}{William K. Black}$^{1}$\thanks{E-mail: wkblack@umich.edu} and
\href{https://orcid.org/0000-0002-4876-956X}{August Evrard}$^{1,2}$
\\
$^{1}$Department of Physics and Leinweber Center for Theoretical Physics, University of Michigan, Ann Arbor, MI 48109, USA\\
$^{2}$Department, Institution, Street Address, City Postal Code, Country
}

\pubyear{2020}

\begin{document}
\label{firstpage}
\pagerange{\pageref{firstpage}--\pageref{lastpage}}
\maketitle


\begin{abstract}
Precision-era optical cluster cosmology calls for a precise definition of the red sequence (RS), consistent across redshift. To this end, we present the Red Dragon algorithm: an error-corrected multivariate Gaussian mixture model (GMM). 
Simultaneous use of multiple colors and smooth evolution of GMM parameters result in a continuous RS and blue cloud (BC) characterization across redshift, avoiding the discontinuities of red fraction inherent in swapping RS selection colors. 
Based on a mid-redshift spectroscopic sample of SDSS galaxies, a RS defined by Red Dragon selects quenched galaxies (low specific star formation rate) with a balanced accuracy of over $90\%$. 
% From the Buzzard flock synthetic galaxy catalogs, we show that a richness calculated by Red Dragon $\lambda_{\rm RD}$ reduces log scatter at fixed mass by $\sim 10\%$ for large halos ($M \gtrsim 10^{14.2}~{M_{\odot}}$) as compared to a fiducial color--magnitude RS selector. 
This approach to galaxy population assignment gives more natural separations between RS and BC galaxies than hard cuts in color--magnitude or color--color spaces. 
The Red Dragon algorithm is publicly available at \href{https://bitbucket.org/wkblack/red-dragon-gamma} {\texttt{bitbucket.org/wkblack/red-dragon-gamma}}. 
\end{abstract}

% Select [1,6] from https://tinyurl.com/ApprovedKeywords

%%%%%%%%%%%%%%%%%%%%%%%%%%%%%%%%%%%%%%%%%%%%%%%%%%
%%%%%%%%%%%%%%%%% BODY OF PAPER %%%%%%%%%%%%%%%%%%
%%%%%%%%%%%%%%%%%%%%%%%%%%%%%%%%%%%%%%%%%%%%%%%%%%

% gus tips: 
% * avoid footnotes
% * avoid overusing acronyms in opening paragraphs

\section{Introduction}
% quick intro 
Galaxies cluster not only in physical space, but in color space as well \citep{Strateva+01,Bell+04}. The advent of CCD technology revealed a strong dichotomy in galaxy colors: a tightly-packed red sequence (RS; predominantly quiescent, passively evolving ellipticals) and a broader blue cloud (BC; predominantly active, star-forming spiral galaxies) \citep{Bower_Lucey_Ellis_1992,Schawinski+14}. Galaxies that fall between the RS and BC populate the `green valley' (GV).
\iffalse 
\footnote{
  Historical note: It was ca. 2007 when the terms BC \& GV gained mainstream traction. \citet{Faber+07} mentions ``blue cloud'' \& `valley' (though it was called a `valley' at least as early as 2004 \citep{Weiner+05}); a few months later, first ``green valley'' references in  \citet{Wyder+07, Schiminovich+07}. 
} \fi 
% The RS is predominantly quiescent, passively evolving ellipticals \citep[$\sim 86\%$; see][]{Schawinski+14} composed of quiescent, passively evolving elliptical galaxies. 
% BC colors follow a broader-peaked $\sim$ Gaussian distribution. The BC is chiefly ($\sim 96\%$) composed of active, star-forming spiral galaxies. 
% The GV contains a mix of the two, with more spirals than ellipticals ($\sim$3::1). 

% causes of the RS
Astrophysically, the RS serves as an imperfect proxy for selecting galaxies with low specific star formation rate (sSFR). 
Star formation decays naturally with age: stellar populations older than roughly $1~{\rm Gyr}$ become almost uniformly red, implying that the reddest galaxies have essentially no star formation \citep{CG10}. 
% Since the galaxy has no luminous young blue stars to maintain the galaxy's hue, it reddens with age. 
% The first galaxies formed in overdensities (of the initial Gaussian-random field that seeded our universe, as seen in the ``baby photo'' of the CMB) that would eventually grow to become the largest halos in our universe \citep{Springel+05}. 
% [Furthermore, since gravity pulls galaxies along large-scale structure towards halos, galaxy clusters tend to accumulate older populations of galaxies.]
Due to the Gaussian random nature of $\Lambda$CDM initial conditions, the density of peaks on different scales are coupled, such that the earliest forming galaxies reside in regions destined to host clusters of galaxies \citep{Springel+05}.  As a result, clusters naturally contain an older galaxy population than the field.  Other dynamical processes that can shut down star formation are also enhanced in proto-cluster environments.
Major mergers between galaxies cause rapid morphological and chromatic shifts from blue spirals towards red ellipticals. Effects such as ram-pressure stripping and AGN feedback blow away gas from high density regions, rapidly diminishing star formation---or {\it quenching}---the galaxy \citep{Schawinski+14}.
Galaxy clusters are natural hotbeds for merging, ram-pressure stripping, and AGN feedback as well, so they serve as ideal nodes at which to find quenched galaxies.

The distribution of sSFR is skew-lognormal, with a peak of blue active star-forming galaxies at ${\rm sSFR} \sim 10^{-10}~{\rm yr}^{-1}$ at low redshift (the galactic main sequence) and a tail towards lower sSFR \citep{Wetzel+12,Eales+18}. This form suggests that the sSFR frequency distribution could be modeled as a dual Gaussian mixture. 
Further strengthening this duality, the scatter in photometric color decreases drastically as sSFR decreases, such that galaxies with ${\rm sSFR} \lesssim 10^{-11.3}~{\rm yr}^{-1}$ share approximately the same color \citep{Eales+17}, thus creating an exceptionally narrow distribution of colors for quiescent galaxies. These factors combined then produce a dual Gaussian in photometric color \citep[see e.g.][]{Baldry+04,Hao+09}: a narrow component for the low-sSFR RS and a wider component for the high-sSFR BC. 

Since the red sequence is particularly strong in clusters, it serves as a strong key for galaxy cluster selection. 
  Identification of clusters by their RS was first proposed by \citet{Gladders&Yee2000}. Since galaxies redden with age, ignoring galaxies bluer than a given cluster's RS removes essentially all galaxies at lower redshifts, efficiently reducing foreground contributions. 
  The maxBCG algorithm \citep{Koester+07} further improved cluster selection, using a hard $\pm 2 \, \sigma_{\rm RS}$ cut in photometric color to select clusters. The algorithm defines richness, $N_{200}$, as the count of red galaxies within an estimated virial radius, $R_{200}$. This count of virialized galaxies within the cluster serves as a halo mass proxy \citep{Rozo+09_constraining}. 
  % The algorithm consequentially finds far more accurate photo-$z$ estimates for clusters than for lone galaxies. % could nix this sentence; I don't deal with photo-z's here. 
  \citet{Rozo+09_improvement} developed an improved richness estimate $\lambda$: the sum of RS membership probabilities for a given cluster, which included a more nuanced cutoff radius and a Gaussian color filter.
  % (fitting color as a function of redshift). 

More recent algorithms and surveys have extended the RS's use for cluster cosmology. 
  To further improve the red/blue galaxy distinction, \citet{Hao+09} developed a single-color error-corrected Gaussian Mixture Model (ECGMM) in color-magnitude space. As compared to a typical GMM, their ECGMM accounted for photometric errors contributing to the scatter. Again, they selected RS galaxies within a hard $\pm 2 \sigma_{\rm RS}$ cut. 
  Around this time, the first results from the SpARCS survey \citep[\SpARCS;][]{Wilson+09,Muzzin+09} produced hundreds of $z>1$ cluster candidates using a selection method similar to that of  \citet{Gladders&Yee2000}. 
  Later, \citet{Rykoff+14} designed the redMaPPer algorithm, which selects RS galaxies in a multi-color + magnitude space, giving a redshift-continuous, multi-color update to richness.  
  Building on similar methodology, \citet{Rozo+16} introduced the redMaGiC algorithm to select luminous red galaxies, estimating galactic redshifts with high accuracy. 
These methods serve as a basis for DES cluster finding in cosmological analyses \citep{Rykoff2016DESSVredmapper, Abbott+20}, with the richness indicator $\lambda$ serving as a mass proxy. 
% Thus the cosmological analyses of e.g. \citet{Abbott+20} used a hard-cut single-color selection of galaxies as their red sequence, 
% \wm{The method we present here is based on a color mixture as \citet{Hao+09}, and acts like RM but w/ more copmonents}

We present {\bf Red Dragon}: a multivariate Gaussian mixture model to select the RS along with other galactic populations. Red Dragon gives a consistent RS definition and continuous red fraction across redshift, characterizing well the underlying photometric distribution of galaxies. 

Red Dragon follows the historical trend of moving from quantized (e.g. binary) classifications towards continuous, probabilistic definitions.
Where once galaxies were purely classified as ellipticals or spirals, % the Hubble tuning fork diagram now contains many sub-groups, spanning from E0 to SBd (\&al. spirals). C
continuous morphological parameters now allow for more precise morphological characterization of galaxies \citep{Conselice_2014}. 
Similarly, the RS has historically been selected as a hard cut in color--magnitude \citep[e.g.][]{Hao+09} or color--color space \citep[e.g.][]{Whitaker+12}, but these cuts lack the nuanced information available from the full multi-color space of 4+ band surveys. 
Red Dragon now offers a probabilistic and smooth RS definition across redshift. 
% Just as YouTube uses clustering in machine learning to target ads to viewers, similar clustering algorithms can now improve the definition of the RS, better distinguishing between RS and BC populations in photometric multi-color space. 

% Outline battleplan here 
We begin in section~\ref{sec:motives} by introducing the datasets used in this analysis, along with an extended discussion of motivations for our method. 
Section~\ref{sec:methods} then details the truth labels used to quantify goodness of RS fit and expounds technical features of the algorithm, such as the optimal number of Gaussian components to use. 
Section~\ref{sec:results} presents several results of this method. 
Finally, section~\ref{sec:conclude} summarizes the method and discusses future applications.

% % % % % % % % % % % % % % % % % % 
% % % % % % % % % % % % % % % % % % 
\section{Data \& Motivations} \label{sec:motives}

% \gus{consider $f_{RS}$ in place of $f_R$ to avoid confusion with modified gravity models.} 

% In addition to smoothing red fraction $f_R(z)$ mean and scatter, several deficiencies of using lower-dimensional color space (i.e. for surveys with $N$ bands, selecting the RS using fewer than $N-1$ colors) suggest the superiority of using redshift-evolving Gaussian mixtures to define the red sequence. 
In this section, we introduce the three datasets used in this work (\S\ref{sec:datasets}) then explain several of our chief motivations in developing this algorithm: chiefly the redshift drift of the 4000~\AA\ break (\S\ref{sec:z_drift}) and the information to be gained from multi-color analysis (\S\ref{sec:beyond}).

% % % % % % % % % % % % % % % % % % 
\subsection{Datasets} \label{sec:datasets}

\begin{table}\centering
  \caption{
    The various datasets used in this analysis. 
    % Each dataset is magnitude-limited to $0.2L_*$, i.e. $m_i < m_*(z) + 1.75$ \citep[using redshift-dependent characteristic magnitude $m_*(z)$ as defined in][]{Rykoff+14}. 
    The baryon pasting algorithm ADDGALS created Buzzard's synthetic galaxy catalog, assigning SDSS-like galaxies to an underlying N-body simulation. 
  }
  \begin{tabular}{ccccc} 
  \hline  
  Dataset & Type & Redshift & sSFR & $N_{\rm gal}$ \\
  \hline 
  SDSS/low-$z$ & Observation & 
    $0.1 \pm 0.005$ & Yes & 44 452 \\ 
  SDSS/mid-$z$ & Observation & 
    $(.3, .5)$ & Yes & 90 609 \\ 
  TNG300-1 & Hydro sim & 
    $0.1$ & Yes & 62 230 \\
  % Buzzard & Mock catalog & $[0.05,0.75]$ & No  & 81 652 780 \\
  Buzzard & Synthetic & 
    $[0.05,0.84]$ & No & 91 004 552 \\
  \hline 
  \end{tabular}
  \label{tab:datasets}
\end{table}

% In this paper, 
We analyze galaxies from the three datasets listed in table~\ref{tab:datasets}: local observed galaxies from SDSS \citep{Szalay+02}, galaxies at low redshift produced by the hydrodynamic simulation IllustrisTNG \citep[][\href{https://www.tng-project.org/data/docs/specifications/\#sec5k}{Model C}, \href{https://www.tng-project.org/files/TNG300-1_StellarPhot/}{observed frame}] {Nelson+18}, and a wide redshift sample from the Buzzard Flock synthetic galaxy catalog \citep{DeRose+19,DeRose+21}. All galaxy samples are luminosity limited such that $L_i(z) > 0.2 \, L_{*,i}(z)$ using the $i$-band characteristic luminosity as a function of redshift, $L_{*,i}(z)$ as defined in \citet{Rykoff+14}. These samples offer complementary tests of Red Dragon's ability to identify the quiescent galaxies of the RS. 

SDSS galaxies were selected from a spectroscopic sample. 
For the low-redshift $z = 0.1 \pm .005$ sample\footnote{SDSS/low-$z$ sample extracted from SDSS \href{http://skyserver.sdss.org/dr16/en/tools/search/sql.aspx}{SkyServer} with \href{https://pastebin.com/HeBmQYqz}{this SQL script}.}, redshift errors were typically $\lesssim 10^{-4}$. We limit summed photometric error to be below $0.3$ to exclude galaxies with poor photometry. Specific star formation rates were calculated using methods from \citet{Conroy+09} and are employed as a truth label to test against Red Dragon's selection of quenched galaxies. 
The other SDSS galaxy sample\footnote{SDSS/mid-$z$ sample extracted from SDSS \href{http://skyserver.sdss.org/dr16/en/tools/search/sql.aspx}{SkyServer} with \href{https://pastebin.com/aAMGY6TW}{this SQL script.}}, spanning redshifts $0.3$ to $0.5$, tests Red Dragon's ability to smoothly select red galaxies as the 4000~\AA\ break crosses filters. This sample has typical redshift error $\lesssim 10^{-3}$, and our required redshift error of less than 0.05 excludes fewer than one in $10^4$ galaxies. 
% These two samples, narrow at low redshift and wide at high redshift, serve distinct purposes. Narrow -> eliminate effects of redshift evolution in fitting populations, low -> more galaxies in the thin bin & less redshift evolution of RS & far from critical redshift. 

We also use the Illustris TNG300-1 cosmological hydrodynamic simulation at redshift $z = 0.1$ (with synthesized SDSS photometry). 
These galaxies have a truth label for sSFR derived from each galaxy's star formation history. As noted in section~\ref{sec:SDSS_limits_of_hard_cuts} and as illustrated in appendix~\ref{apx:SDSS_vs_TNG_hist}, the distributions of galaxy colors do not match well those of SDSS, making application to this sample more so a test of Red Dragon's robustness than of its RS selection capacity. 

% Even after adding SDSS-like photometric errors, the sample still has a far stronger GV than SDSS (see appendix~\ref{apx:SDSS_vs_TNG_hist}), so we focus our analyses here on SDSS and Buzzard. Our fitting of TNG does, however, show the robustness of Gaussian mixture models in fitting various distributions of galaxy photometries. 

The Buzzard synthetic galaxy catalog is a wide-area galaxy sample that extends the redshift range of our analysis to $z = 0.84$.
To create the Buzzard galaxy catalog, the {\sc ADDGALS} algorithm \citep{BW08,Wechsler+21} populated lightcone outputs of N-body simulations with galaxies.  The empirical method introduces galaxy bias using a local dark matter density measure, and colors are applied using templates tuned to SDSS and other observed galaxy samples.  The method reproduces well the magntitude counts and two-point clustering of galaxies at $z < 1$, but massive clusters are somewhat underpopulated as compared to observations \citep{DeRose+21, Wechsler+21}. 
% This makes it a nice toolbox for fitting galaxy photometry.
% We limit our consideration to only galaxies with $m_i < Only galaxies with m_*(z) + 1.75$ \citep[using redshift-dependent characteristic magnitude $m_*(z)$ as defined in][, equation (9)]{Rykoff+14}. 

% % % % % % % % % % % % % % % % % % % % % 

\begin{figure*}\centering
  \includegraphics [width=\linewidth] {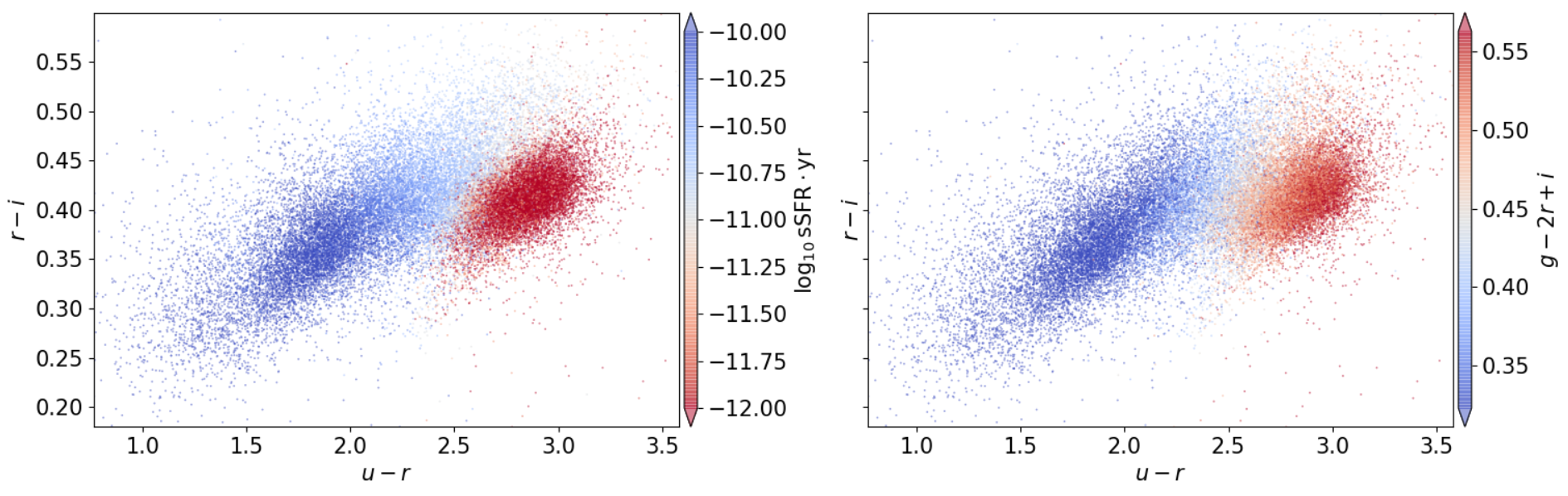}
  \caption{
    SDSS low-redshift galaxy sample ($z = 0.1 \pm 0.005$) plotted in CC space. 
    The color $u-r$ gives a relatively clean measurement of the strength of the 4000~\AA\ break, measuring current SFR. 
    The color $r-i$ measures the post-break slope, an indicator of dust content (somewhat degenerate with age and metallicity). 
    {\bf Left:} colored by log specific star formation rate, with values below roughly -11 corresponding to the quenched population. 
    {\bf Right:} colored by $g -2r +i$, a pseudo finite difference second derivative near the 4k\AA\ break, corresponding roughly with metallicity. 
  }
  \label{fig:bi_panel}
\end{figure*}

% % % % % % % % % % % % % % % % % % % % % 
% redshift drift requiring more bands

\subsection{Redshift drift of 4000~\AA\ break} \label{sec:z_drift}

% This subsection discusses why the redshift drift of the 4000~\AA\ break suggests a superiority of a multi-color model (using all relevant colors simultaneously). 

Our main motivation in creating the Red Dragon algorithm originated in the redshift drift of 
% Spectroscopically, the quenching of a galaxy manifests primarily in 
``the 4000~\AA\ break'': a sharp drop in spectral intensity at short wavelengths and the primary effect of quenching on a galaxy's spectrum. 
This break has two main sources. 
% The shutdown of star formation means no new bright blue stars form; existing bright blue stars quickly redden with age.
Though only certain wavelengths larger than 3645~\AA\ can be absorbed by excited ($n=2$) Hydrogen, any wavelength shorter than that will fully ionize an electron. This asymptote of the Balmer series at 3645~\AA\ results in a sharp drop of intensity towards shorter wavelengths \citep{Mihalas_1967}. 
Meanwhile, stellar production of metals results in a blanket of line absorption, reddening the spectrum around 4000~\AA~\citep{Worthey_1994}. 
These two effects conspire to cause a strong suppression of emission at wavelengths shorter than $3800 \pm 200$~\AA. % the break. 

% Table of redshift for 4kA break
\begin{table}\centering
  \caption{ % Remember to define the quantities, symbols and units used.
    Approximate redshift ranges over which each band will measure 4k\AA\ rest wavelengths, given for both \href{https://www.sdss.org/instruments/camera/\#Filters}{SDSS} \citep{Doi+10} and \href{http://data.darkenergysurvey.org/aux/releasenotes/DESDMrelease.html}{DES} \citep{Abbott+18} photometries. % Note the significant overlap of z and Y bands. 
    % Values give the approximate minimum redshift at which the break's presence would be noticed in this band. 
    % Errors are approximately $\Delta z \sim \mathcal{O}(0.1)$ redwards, since the jump spans down to the Balmer break, at 3645~\AA. 
    % As a mnemonic, the bands roughly correspond to light coming from ultraviolet (u), green (g), red (r), near infrared (i), and longer wavelengths up to a micrometer (z, Y). 
    % in SDSS photometry, they get the most response out of g,r,i---u and z are pretty small in comparison. 
  }
  \label{tab:4kA_break}
  \begin{tabular}{lll} % four columns, alignment for each
    \hline 
	band & $z_{\rm break,SDSS}$ & $z_{\rm break,DES}$ \\
	\hline 
% 	u & $[-0.2,  0.0)$  & N/A \\
	g & $[ 0.0,  0.36)$ & $[ 0.0,  0.38)$ \\
	r & $[ 0.36, 0.71)$ & $[ 0.38, 0.78)$ \\
	i & $[ 0.71, 1.06)$ & $[ 0.78, 1.13)$ \\
	z & $[ 1.06, 1.37)$ & $[ 1.13, 1.50)$ \\ % 1.515 for 2% of max transmission 
	Y & N/A             & $[ 1.38, 1.55)$ \\
	\hline 
  \end{tabular} 
\end{table}

Table~\ref{tab:4kA_break} shows approximate redshifts at which a rest frame wavelength of 4000~\AA\ is observed in each observational band for both SDSS and DES. The difference in magnitude of the bands {\it surrounding} the band in which the break resides gives the cleanest measure of D4000 (the ratio of intensity on either side of the break), which gives an excellent estimate of sSFR. The optimal photometric color for RS selection thus changes with redshift. 
% (Figure~\ref{fig:SDSS_Z} shows the waxing and waning of RS selection power for individual bands across redshift). 

If a RS selector uses only one color at a time, with discrete jumps in photometric color at certain transition redshifts, these hard transitions can result in an $\mathcal{O}(10\%)$ shift in red fraction, $f_R(z)$ \citep[up to $\sim 16\%$; see e.g.][]{Nishizawa+18}. 
% low-redshift jumps from Nishizawa+18 Fig~7: 
% >>> [0.117, 0.157, 0.133] = 14% +/- 2%
% fR jumps from SAM: 
% >>> [0.055, 0.047, 0.041, 0.043, 0.026, 0.031] = 4.0% +/- 1%
This jolt in red fraction would echo in single-color richness estimates based on a count of bright red galaxies in a cluster. Two identical clusters on either side of a redshift transition could then have significantly different $\lambda_{\rm col}$ values, introducing non-trivial systematic errors in halo mass--richness scaling relations.

Evolving a multi-color Gaussian mixture across redshift smoothly defines the RS, obviating the discontinuities caused by color swapping. 
Taking all colors into account simultaneously allows for a continuous and consistent RS out to high redshifts.

% % % % % % % % % % % % % % % % % % % % % 
% missing out on features beyond quenched status 

\subsection{Beyond the 4000~\AA\ break} \label{sec:beyond}

Though a galaxy's quenched status primarily manifests though the strength of D4000, %(the ratio of flux on either side of the 4000~\AA\ break), % don't need this again so soon; it's not the focus here anyway. 
other astrophysical factors such as age, dust, or metallicity separate RS from BC photometrically. 
(For a summary of main effects of galaxy properties on optical spectra, see Figure~\ref{fig:cartoon_gal_spec}.) 
Though D4000 can be estimated using a single photometric color, multi-color analysis serves to better distinguish the RS from the BC. 

% How various methods did at selecting things
% 
% Slopes (single C): 
% u-r → 	91%
% g-r → 	88.5%
% r-i → 	66%
% i-z → 	77.3% 
% 
% Curvatures: 
% u-2g+r → 85.1%
% g-2r+i → 93.1%
% r-2i+z → 75.5%

Figure~\ref{fig:bi_panel} illustrates this for the low-$z$ SDSS galaxy sample. While the horizontal axis $(u-r)$ correlates highly with D4000, the vertical axis $(r-i)$ gives a degenerate measure of dust content and other properties. The left panel colors points by specific star-formation rate (where $\log_{10} {\rm sSFR} \sim -11$ separates quenched from star-forming galaxies) while the right panel colors points by $(g-2r+i)$. The latter visibly correlates with the former. 

This composite feature of $(g - 2r + i)$ acts as a pseudo second derivative, approximating here the spectrum curvature near the 4000~\AA\ break. While ordinary single-color (a vertical line on this plot) or CC selection (an angled line on this plot) would be ignorant of such information, the curvature information clearly correlates with sSFR and would aid in selecting the quenched population. Even a perfectly positioned hard line cut would be inherently limited in selecting quenched galaxies (see Figure~\ref{fig:bACC_CM_CC_GM}). 

A multi-color Gaussian mixture simply includes such curvature terms using the primary color space (i.e. differences between neighboring bands; see equation~\ref{eqn:primary_SDSS}). Here we have $(u-r) = (u-g) + (g-r)$ and $(g - 2r + i) = (g-r) - (r-i)$, showing up as $\pm 45^\circ$ directions in the multi-dimensional primary color space. 
Furthermore, the populations overlap in both CM and CC spaces, limiting the power of hard cut selection. In contrast, Gaussian mixtures are {\it designed} to model such overlapping populations, making them a natural tool to consider in selecting the RS and BC.

In order to combat $f_R(z)$ discontinuities, move beyond hard cuts in photometry, and better select the photometry-space population of RS galaxies, we present the Red Dragon algorithm.

%%%%%%%%%%%%%%%%%%%%%%%%%%%%%%%%%%%%
%%%%%%%%%%%%%%%%%%%%%%%%%%%%%%%%%%%%
\section{Methods} \label{sec:methods}

Red Dragon is a novel method for calculating red sequence membership probabilities $P_{\rm RS}$. In its most general construction, a Red Dragon RS selector uses a Gaussian mixture in multi-color space to select populations of galaxies (RS, BC, and optionally additional components). 
% After calculating mixture properties in discrete redshift shells, the GMM parameters are interpolated across redshift, resulting in a smooth and consistent definition of the RS. 
In \S\ref{sec:construction}, we outline the algorithm, including the sequence of operations and the relevant likelihood function. 
Considerations when applying the algorithm are presented in \S\ref{sec:considerations}, discussing choices such as the optimal number of colors or model components.

% % % % % % % % % % % % % % % % % % 
% % % % % % % % % % % % % % % % % % 
\subsection{Algorithm Construction} \label{sec:construction}
Here we give an overview of the algorithm (\S\ref{sec:overview}), introduce the core likelihood function for Red Dragon (\S\ref{sec:llh}), and detail interpolation of GMM parameters across redshift (\S\ref{sec:interpolation}).

% % % % % % % % % % % % % % % % % % 
\subsubsection{Overview of algorithm} \label{sec:overview}

\begin{figure}\centering
  \begin{tikzpicture}[node distance=1.5cm]
    % % % % % % % % % % % % % % % % % % 
    % set up RD DNA process
    \node (input) [io] {Input Data: \\ $z$, $\delta_z$; $\vec{m}$, $\vec{\delta}_m$};
    \node (lbl1) [vertex, left of=input, xshift=-1cm] {1}; 
    \node (z_sel) [process, below of=input] {Probabilistic Redshift Selection};
    \node (lbl2) [vertex, left of=z_sel, xshift=-1cm] {2}; 
    \node (GMM) [process, below of=z_sel] {GMM};
    \node (lbl3) [vertex, left of=GMM, xshift=-1cm] {3}; 
    \node (output) [io, below of=GMM] {Output~Data: $\{ \vec{\theta}_\alpha \}$ (per~$z$-bin)};
    \node (lbl4) [vertex, left of=output, xshift=-1cm] {4}; 
    % set up arrows
    \draw [arrow] (input) -- (z_sel);
    \draw [arrow] (z_sel) -- (GMM);
    \draw [arrow] (GMM) -- (output);
    % % % % % % % % % % % % % % % % % % 
    % set up RD interpolation process
    % \node (theta) [io, xshift=3cm, right of=input] {$\{ \vec{\theta} \}$ at each $z$-bin};
    % \node (nan) [process, below of=theta] {Check for nan values \& exclude};
    \node (match) [process, below of=output] {Component Matching}; 
    \node (lbl5) [vertex, left of=match, xshift=-1cm] {5}; 
    % \node (outlier) [process, below of=match, yshift=-.5cm] {Outliers Exclusion}; 
    \node (interp) [process, below of=match] {Interpolate}; 
    \node (lbl6) [vertex, left of=interp, xshift=-1cm] {6}; 
    \node (RD) [io, below of=interp] {Red Dragon: $\vec\theta_\alpha(z)$, $P_{\rm mem}$}; 
    \node (lbl7) [vertex, left of=RD, xshift=-1cm] {7}; 
    % set up arrows
    \draw [arrow] (output) -- (match); 
    \draw [arrow] (match) -- (interp); 
    \draw [arrow] (interp) -- (RD);
    % try to draw arrow back to GMM
    % \draw [arrow] (RD) -| (2.5,-3) -- (GMM); 
    \draw [arrow,gray] (RD) -| ([xshift=.75cm]RD.east) |-  node[anchor=south,rotate=270,xshift=3cm]{re-initialize components} (GMM);
    \node (lbl8) [vertex, right of=match, xshift=+1.5cm] {8}; 
  \end{tikzpicture}
  \caption{
    Work flow for RD algorithm. The process has two main portions: (1--4) moving from input data to creating GMM parameterizations $\{ \vec\theta_\alpha \}_i$ at each redshift slice $i$ and (4--7) interlinking components across redshift from the disarranged collection $\{ \vec\theta_\alpha \}$ to parameterize redshift-continuous GMM components, $\vec\theta_\alpha(z)$, and thereby membership probabilities $P_{{\rm mem},\alpha}(z,\vec m, \vec \sigma_m)$. 
    % dashed arrow
    The grey arrow (8) indicates that after a dragon is created, it may be used to initialize components in fitting the GMM. 
    This is especially useful for outlier redshift bins which didn't fit like their neighbors. 
    By default, the algorithm does a single pass (1--7) fitting sparsely with the \texttt{sklearn} GMM, then uses that fit to inform initial conditions for the \texttt{pyGMMis} GMM, which gives the second pass (4--7) for a final fitting. 
  }
  \label{fig:flowchart}
\end{figure}
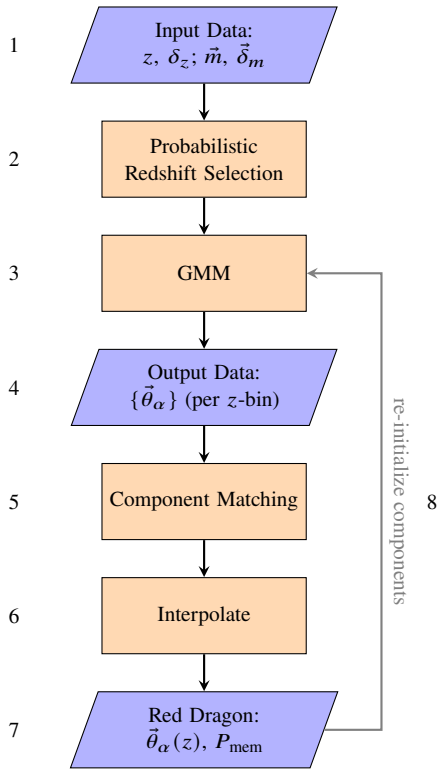

Broadly speaking, there are two stages to the Red Dragon algorithm, as illustrated in Figure~\ref{fig:flowchart}. 
In the first stage, it segments the data into discrete redshift shells\footnote{
  Though we describe fits as functions of redshift here, any secondary variable may be used (given a thin enough redshift extent for the data), such as stellar mass or a single photometric band. 
} and finds GMM parameterizations for each slice.
In the second stage, it matches components across redshift bins and interpolates parameters. This results in continuous and consistent definition of the RS and BC (and optionally further components).

\paragraph*{First stage: redshift-discrete fits.} 
Red Dragon reads in photometry and redshift information as supplied by the user (see the values in Figure~\ref{fig:flowchart}, box~1). 
Input redshift estimates $z$ and errors $\delta_z$ allow for binned analysis of redshift evolution. 
From input magnitudes $\vec m$ and magnitude errors $\vec\delta_m$, Red Dragon calculates colors $\vec c$ and the corresponding noise covariance matrix $\Delta$ (see equation~\ref{eqn:noise}). 
% Redshift estimates $z$ and errors $\sigma_z$ allow for analysis of the redshift evolution of the fit.  \gus{Where do redshifts come from?  Aren't they essential for this step?}  Alternatively, for thin redshift bins, one may replace the `redshift' variable with magnitude (or other secondary features, such as local density) \gus{Stick to z here, you've already mentioned the other flavors of analysis above.}. 
This set of input variables are then sent to a GMM to find fit parameters $\vec \theta_\alpha$ in each redshift bin. See section~\ref{sec:llh} for details on the likelihood model and fitting process.

At this point, the algorithm has saved as output data $\{ \vec\theta_\alpha \}_i$, i.e. at each redshift bin $i$ and for each component $\alpha$, it saves Gaussian mixture parameters $\vec\theta$ (see table~\ref{tab:parameters}). 
These components are unordered at this point, so the $\alpha=0$ component in redshift bin $i=1$ may not correspond to the $\alpha=0$ component in redshift bin $i=2$. 
% Each redshift bin has its own disarranged % unsystematic / disarranged / disarrayed / slipshod / disordered 
% collection of Gaussians, characterizing the populations of each slice. 
% 
% serves as the ``DNA'' of the dragon. It still must be pieced together and tamed---that is---the 
This set of Gaussian parameterizations must be linked across redshift continuously to avoid spurious rapid changes of classification over small redshift ranges.

\paragraph*{Second stage: redshift-continuous fits.}
Using the calculated set of Gaussian mixtures, $\{ \vec\theta_\alpha \}_i$, the algorithm now matches similar Gaussian components across redshift bins. 
While distinguishing continuous components across redshift is relatively simple for two-component models ($K=2$), matching for even three components ($K=3$) can be challenging. 
  (Does the wide, redder portion of the BC connect to the narrower but similar in color component in the adjacent redshift bin, or does it connect to the component with similar scatter despite its significantly bluer color?) 
Despite these challenges, the matching process can be largely automated, as discussed in appendix~\ref{apx:continuity}. 

After successful matching across redshift bins, the now linked set of parameters $\{ \vec\theta_{\alpha'} \}_i$ can then be interpolated across redshift, giving the continuous parameterizations $\vec\theta_\alpha(z)$. These continuously evolving Gaussian components can then yield component membership probabilities for a galaxy at a given redshift with given photometry for each component $\alpha$: $P_{\rm mem,\alpha} (z, \vec{m}, \vec{\delta}_m)$. 
% These can, of course, then be discretized to give boolean membership assignment, but this loses power, as compared to continuous membership probabilities. [not sure I have quantitative backing for this other than generalizations of "more info = better"]
This interpolated parameterization now presents the user with a trained dragon, smoothly characterizing each of the populations across redshift.

% % % % % % % % % % % % % % % % % % 
\subsubsection{Likelihood model} \label{sec:llh}

\begin{table}\centering
  \begin{tabular}{clcl}
    \hline
    $\theta_\alpha$ & model parameters &
    $x_j$ & galaxy data  \\
    \hline 
    $w_\alpha$ & weight (where $\sum_\alpha w_\alpha = 1$) &
    $\vec c_j$ & color \\
    $\vec \mu_\alpha$ & mean color &
    $\vec \delta_j$ & errors on galaxy colors \\
    $\Sigma_\alpha$ & intrinsic covariance &
    $\Delta_j$ & noise covariance \\
    \hline 
  \end{tabular}
  \caption{
    Summary of variables in likelihood model (see equations~\eqref{eqn:L_ij} and~\eqref{eqn:RD6C}). 
    Left column shows GMM model parameters $\theta_\alpha$ characterizing component $\alpha$ of $K$; right column shows input data for galaxy $j$ of $N_{\rm gal}$. 
    Galaxy colors, with their respective errors and noise covariance, are calculated from input magnitudes $\vec m \pm \vec\delta_m$ within redshift bins (determined by galaxy $z \pm \delta_z$). 
  }
  \label{tab:parameters}
\end{table}

Red Dragon employs a multi-dimensional Gaussian mixture model which accounts for photometric errors; parameters of the Gaussian mixture evolve with redshift to give continuous characterizations of each GMM component. 
%\gus{Stick to the basics here: the secondary parameter is redshift, period.} 
% Though the running of Gaussian parameters with magnitude causes minimal selection differences (see section~\ref{sec:magnitude_run}), the running of Gaussian parameters with redshift is sharply non-linear across wide spans in redshift spans.
% and mandate accommodation. 
% (or in practice, across any other secondary parameter given a sufficiently thin redshift slice). See appendix~\ref{apx:quan_mag_run} for more details regarding evolution with secondary parameters. 

A Gaussian mixture of $K$ components in a single color $c$ has a set of parameters $\theta$ that constitute the model. For each component $\alpha$, this parameter set includes the component weight $w_\alpha$, mean color $\mu_\alpha$, and the intrinsic population scatter $\sigma_\alpha$ (see table~\ref{tab:parameters} for a summary of parameters used). The weights are normalized such that $\sum_\alpha w_\alpha = 1$. For a set of $N_{\rm gal}$ galaxies with input colors $\{ c_j \}$, and color errors $\{ \delta_i \}$, the model parameters maximize the likelihood 
\begin{equation} \label{eqn:H09}
  \mathcal{L} ( \theta \big| x ) = \prod_{j=1}^{N_{\rm gal}} \left\{ \sum_{\alpha=1}^K \frac{w_\alpha}{\sqrt{2\pi ({\sigma_\alpha}^2 + {\delta_j}^2)}} \exp\left[ - \frac12 \frac{(c_j-\mu_\alpha)^2}{{\sigma_\alpha}^2 + {\delta_j}^2} \right] \right\} .
\end{equation}
This type of error-corrected Gaussian mixture model (ECGMM) was introduced by \citet{Hao+09} with SDSS $g-r$ as the color classifier.

% variation with magnitude cf redshift: 
% fR >> 'gmr_BC' > 'sig_BC' = 'sig_RS' > 'gmr_RS' 

% significantly non-zero slopes: 
% fR > sig_RS > gmr_BC > gmr_RS > sig_BC

Expanding this model into an $N$-dimensional color space requires that we employ for each component $\alpha$ an intrinsic color covariance matrix $\Sigma_\alpha$. The errors then must be handled as a noise covariance matrix $\Delta_i$ for each galaxy. 
% \gus{Maybe we should use bold $\mathbf{\Sigma}$ and  $\mathbf{\Delta}$?}  

% The noise covariance matrix $\Delta$ has analytic calculation from photometric errors. 
% If some linear relation $f$ transforms $\vec x$ into $\vec y = f(\vec x) = A \, \vec x$, then the noise covariance $\Delta_x$ is transformed accordingly as $\Delta_y = A \, \Delta_x \, A^\top$. 

Consider the DES four-band optical $griz$ photometry used in Buzzard, with input magnitudes $\vec m = [m_g, \, m_r, \, m_i, \, m_z]$.  We define a vector of {\bf primary colors} based on neighboring photometric bands: 
% using shorter-wavelength bands as minuends. 
% For DES photometry, we have
\begin{equation}\label{eqn:primarycolor}
  \vec c = [g-r, \, r-i, \, i-z]. 
\end{equation}
Colors are derived from magnitudes by the matrix operation $\vec c = A \, \vec m$, where the transform matrix is
\begin{equation}
  A = \begin{bmatrix}
      1 & -1 & 0 & 0 \\
      0 & 1 & -1 & 0 \\
      0 & 0 & 1 & -1
  \end{bmatrix}.
\end{equation}
We assume that the photometric errors of each galaxy are determined independently in each band, and so take them to be uncorrelated. The magnitude error covariance matrix $M_j$ for each galaxy is then diagonal. 
% 
% Using the relation from above, 
Transformed to the space of primary colors (equation~\eqref{eqn:primarycolor}), the noise covariance of galaxy $j$ is then
% \gus{But $\sigma$'s are the intrinsic population scatter, right?  Don't you mean to use $\delta$'s here?} 
% \wm{these are bands, not colors, so technically as written, e.g. $\delta_{(g-r)} = \sigma_g^2 + \sigma_r^2$, but the notation here is certainly convoluted...}
\begin{equation} \label{eqn:noise}
  \Delta_j = A \, M_j \, A^\top  =
  \begin{bmatrix}
    {\delta_g}^2 + {\delta_r}^2 & -{\delta_r}^2 & 0 \\
    -{\delta_r}^2 & {\delta_r}^2 + {\delta_i}^2 & -{\delta_i}^2 \\
    0 & -{\delta_i}^2 & {\delta_i}^2 + {\delta_z}^2
  \end{bmatrix}_j
\end{equation}
where $\delta_x$ above refer to the photometric error of band~$x$ for galaxy $j$. % the subscripts above refer to band. 
Note that for this matrix to be non-singular, the selection of colors must be linearly independent (e.g. one cannot use each of $g-r$, $r-i$, and $g-i$ in an error-inclusive model). For symmetry, simplicity, 
% (keeping colors from spanning many bands, like $g-z$), we
and to avoid singularity, we employ the set of primary colors. 

The derivation of $\Delta_j$ is similar for SDSS photometry (which includes $u$ band). The primary color vector for $ugriz$ is then
\begin{equation}\label{eqn:primary_SDSS}
  \vec c = [u-g, \, g-r, \, r-i, \, i-z]
\end{equation}
and the corresponding $\Delta_j$ matrices come from a straightforward extension of the above matrices. % of $A$ and $\Delta_j$. 

This likelihood of this error-cognizant $N$-dimensional Gaussian mixture model is then
\begin{equation} \label{eqn:RD6C}
  \mathcal{L} ( \theta \big| x) = 
  \prod_{j=1}^{N_{\rm gal}} \sum_{\alpha=1}^K
    \mathcal{L}_\alpha ( \theta_\alpha \big| x_j)
\end{equation} 
where the likelihood for each galaxy $j$, component $\alpha$, is
\begin{equation}
\begin{aligned} \label{eqn:L_ij}
  \mathcal{L}_\alpha ( \theta_\alpha \big| x_j) = \, 
  & \frac{w_\alpha}{\sqrt{(2\pi)^N} \left| \Sigma_\alpha + \Delta_j \right|} \\
    & \times \exp 
      \left[ 
        -\frac12 
        (\vec c_j - \vec \mu_\alpha)^\mathrm{T} 
        (\Sigma_\alpha + \Delta_j)^{-1}
        (\vec c_j - \vec\mu_\alpha) 
      \right] .
\end{aligned}
\end{equation}
% for component $\alpha$ and galaxy $j$, 
Here, $x_j$ includes all primary colors $\vec c_j$ as well as the noise covariance matrix $\Delta_j$ for each galaxy (see table~\ref{tab:parameters}).

At individual redshift slices, we use the error-inclusive Gaussian Mixture package \href{https://github.com/pmelchior/pygmmis} {pyGMMis} \citep{Melchior_Goulding_2018} to find best-fit parameters $\theta_\alpha$ for each component $\alpha$.
Without a reasonable input for a first guess at parameters, pyGMMis sometimes struggles to properly characterize populations. To provide a rough first guess, we first sparsely fit the data using \texttt{sklearn}'s error ignorant \href{https://scikit-learn.org/stable/modules/generated/sklearn.mixture.GaussianMixture.html}{GaussianMixture} package \citep{sklearn}. 
This extremely quick fit gives a rough initial guess to the fit parameters, yielding better results than running \texttt{pyGMMis} blind.

% % % % % % % % % % % % % % % % % % 
\subsubsection{Fit interpolation} \label{sec:interpolation}
Red Dragon interpolates best-fit parameters across redshift bins, continuously defining populations. 
After fitting weights, the normalization $\sum_\alpha w_\alpha(z) = 1$ is re-enforced. % How?
To interpolate the covariance matrix, log variances are interpolated first, followed by interpolating the correlations (enforcing $|\rho| \leq 1$), which together then provide a better fit than purely fitting the covariance matrix all at once (which could result in unphysical negative variances). 
% \wm{Take care in training dragons; improper interpolation could result in negative variances or unrealistic end behaviors.} 
Fitting is linear by default (with flat endpoint extrapolation), but other methods such as smoothed spline interpolation \citep[SciPy:][]{2020SciPy-NMeth} or kernel-localized linear regression \citep[KLLR:][] {Farahi+18, Anbajagane+20} are available to give smoother fits. 

% Once a Red Dragon has its DNA coded from the initial $\Delta z$ bin analysis, 
These redshift-continuous fits can then predict for individual galaxies its membership likelihood for each component. 
% From the individual component likelihoods of equation~\ref{eqn:RD6C}, one can calculate membership probabilities. In general, 
The probability that galaxy $j$ is a member of GMM component $\alpha$ is 
\begin{equation}
  P_\alpha(x_j) = \frac{ \mathcal{L}_\alpha (\theta_\alpha \big| x_j) } { \sum_\beta \mathcal{L}_\beta ( \theta_\beta \big| x_j)}. 
\end{equation}
A two-component model would then have red sequence membership probability $P_{\rm RS} = \mathcal{L}_{\rm RS} / (\mathcal{L}_{\rm RS} + \mathcal{L}_{\rm BC})$. 

% Note that in selecting a slice, photometric redshift errors can be accounted for by integrating the likelihood of equation~\ref{eqn:RD6C} across the distribution of redshift estimates $P(z_{\rm photo})$, weighted by relative likelihood of each redshift. In our algorithm, we allow probabilistic selection of galaxies for each redshift bin. 

This parameterization results in a redshift-continuous definition of the red sequence over large redshift spans, without the jumps or transitions incurred by single or double-color RS selection. Its more objective definition of the RS better characterizes the nuances of galaxy multi-color space than hard cuts. % \wm{[maybe a bit too strong]} 

% % % % % % % % % % % % % % % % % % 
% % % % % % % % % % % % % % % % % % 
\subsection{Algorithm Considerations} \label{sec:considerations}
In this section, we 
  define accuracy in selecting the quenched population for this analysis (\S\ref{sec:bACC}), 
  detail the accuracy gains from added bands (\S\ref{sec:optimal_bands}), 
  discuss the optimal count of Gaussian components (\S\ref{sec:elment_count}), 
  and discuss whether Gaussian features must be allowed to run with magnitude to accurately select the quiescent population (\S\ref{sec:magnitude_run}).

% % % % % % % % % % % % % % % % % % 
\subsubsection{Balanced Accuracy} \label{sec:bACC}

To quantify goodness of fit for the RS, we use the binary classification measure of `balanced accuracy,' comparing RS members selected by Red Dragon to the quenched population.  We convert Red Dragon red component probability to a binary RS classifier by the condition $P_{\rm RS} > 0.5$ and defined quenched galaxies using a threshold in specific star formation rate as a function of redshift 
\begin{equation}\label{eqn:quenched}
  \log_{10} \left( {\rm sSFR} \cdot {\rm yr} \right) < -11 + z
\end{equation}
\citep[adapted from][for our mass and redshift ranges]{Moustakas+13}. 
A more complicated determination of a truth label for the RS could include measures of stellar mass, dust, metallicity, and age as metrics to aid in separating RS and BC (see section~\ref{sec:beyond}). For example, one could simply add these into a GM or other machine learning structure along with the colors, giving the structure more information to aid in the separation. 
While such a model may serve as a more accurate truth label to test against, our benchmark hard cut in sSFR defines a straightforward underlying truth in the photometric distribution of galaxies; its strong correlation with idealized galaxy characterization gives it value in discriminating between RS selectors.

Balanced Accuracy (also BA or bACC) takes the average of sensitivity and specificity, i.e. the true positive rate TPR$\equiv$TP/(TP+FN) and the true negative rate TNR$\equiv$TN/(TN+FP). 
This compensates for unequal population ratios: 
 the relative weight between RS and BC varies significantly across redshift and magnitude, so bACC equally represents selection accuracy between the two populations.\footnote{
  Note that a score of 50\% would be earned by a worthless test categorizing all as either solely positive or negative. %---such a worthless test deserves a score of zero, which is good reason to prefer Youden's $J$ statistic (equal to $2{\rm bACC}-1$). However, the obscure name could have distracted from the message at hand, hence our using bACC.
}

% \wm{[Include \P\ like the following?]:} 
We caution the reader that achieving 100\% accuracy is not only practically impossible (without overfitting), but also not quite ideal. Since the sSFR distributions of RS and BC overlap, a hard cut in sSFR to score selection would mischaracterize a set fraction of galaxies from each. More nuanced selection of the RS and BC (defined from the more complicated definition above) would then have an accuracy at some value below 100\%, though still high. Therefore, balanced accuracies from a hard cut in sSFR below 100\% should be no cause for worry, and indeed, could indicate the method is working properly. 

% concession answering question along the lines of "why are we using a hard cut if we're saying hard cuts are a thing of the past?" 
Balanced accuracy requires a binary classification, so it is somewhat limited in its ability to score goodness of fit for ambiguous cases where e.g. the quen\-ch\-ed probability $P_{\rm Q} = 49\%$ (the chance, taking error bars into account, that equation~\ref{eqn:quenched} is true) but the RS membership probability $P_{\rm RS} = 51\%$ (the chance, derived from Gaussian Mixtures, that it belongs to the redder component). 
The sSFR distribution is lognormal skewed, with the bulk of galaxies falling near the sSFR cut of equation~\eqref{eqn:quenched}, so middle probabilities are common, with $\sim 40\%$ of galaxies lying in $P_Q|(25,75)\%$. Quenched probabilities are therefore somewhat sensitive to the sSFR cut one uses to define the quenched population; any hard cut in sSFR will necessarily change the resulting bACC. 
However, the distribution of $P_{\rm RS}$ values is strongly bimodal, with generally $\lesssim 15\%$ of galaxies lying in $P_{\rm RS}|(25,75)\%$ for Buzzard and $\lesssim 1\%$ for SDSS (TNG sSFR values are without errors). 
Since probabilities generated by Red Dragon tend towards zero and one, the problem of hard cuts in binary classification is somewhat mitigated. 
A simple binary classification metric aptly characterizes the large majority of galaxies and gives a simple measure for RS selection power. 
% Furthermore, it lends itself to easier interpretation than more complicated methods, so we stick to bACC for this analysis. 

% % % % % % % % % % % % % % % % % % 
\subsubsection{Accuracy gains from added colors} \label{sec:optimal_bands}

Typically, single-color RS selection uses a color constructed from the photometric band containing the 4000~\AA\ break and the longer-wavelength band immediately after ($g-r$ at low redshifts); two-color RS selection further includes the primary color with the next longest wavelength band after that ($r-i$ at low redshifts). However, other colors can aid in better distinguishing the RS from the BC. 

For SDSS low-$z$, we used all possible colors (including the band-jumping secondary colors, like $g-i$ or $u-z$, in addition to the primary colors; only considering non-singular combinations of colors) to create single, double, and triple color Gaussian mixture models, revealing optimal color combinations along with accuracy gains from adding colors. 
Comparing these optimized color groupings to our choice of using all primary colors for Red Dragon's spine, we can gauge to what extent selection accuracy depends on choice of the input color vector $\vec c$. 

\begin{figure}\centering
  \includegraphics [width=\linewidth] {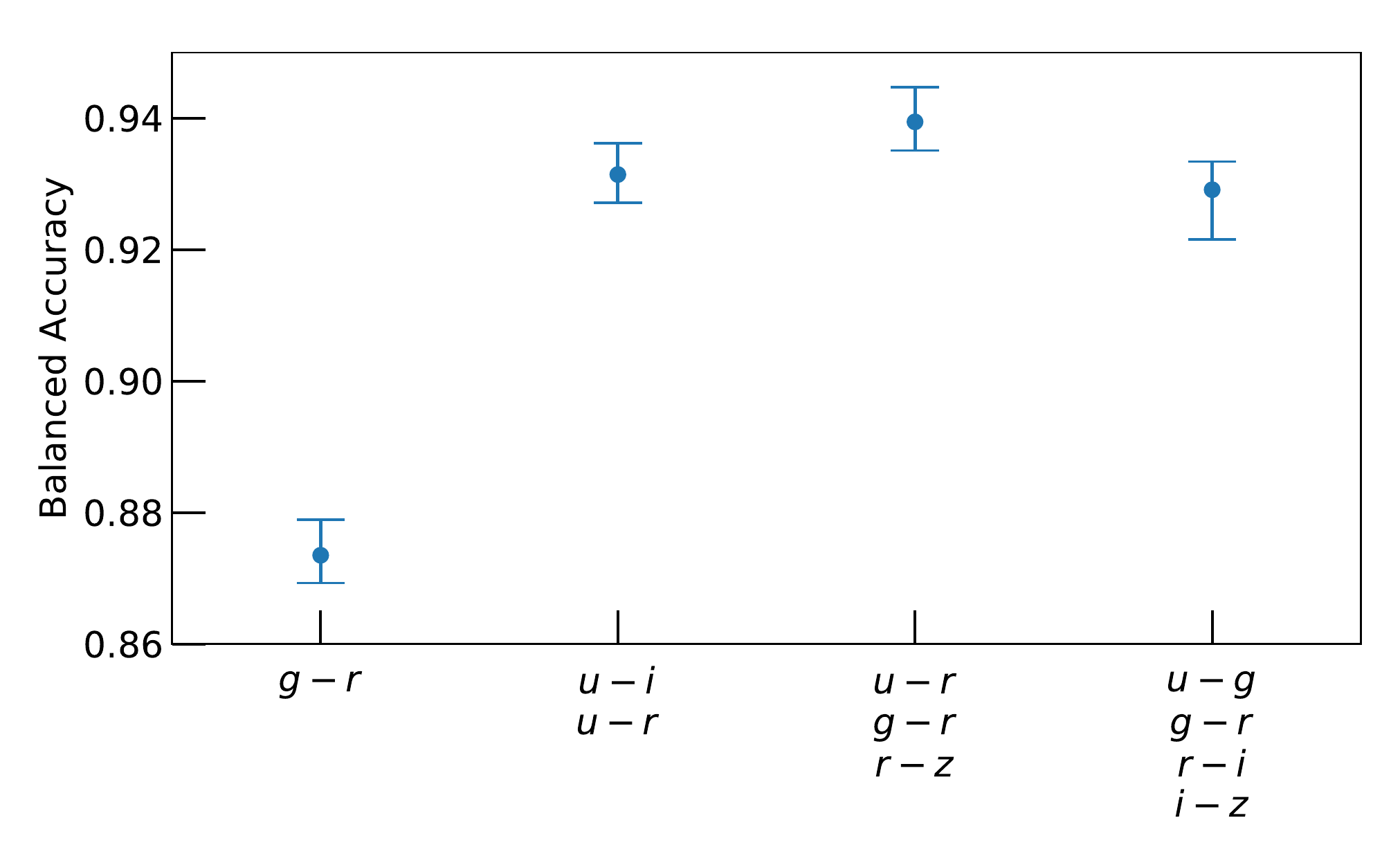}
  \vspace{-.75cm}
  \caption{
    Accuracy of selecting the quenched population in low-$z$ SDSS by Gaussian mixture for various combinations of input colors.
    Each of the first three $N$-color groupings were optimized in their selection, having the highest median bACC of all possible $N$-color combinations. 
    Error bar plot shows bootstrap $\pm 2 \sigma$ quantiles. 
    % 
    % Note that due to using linearly dependent combinations of colors (extending beyond the primary color vector $\vec c_4$ which spans the full color space), the $\vec c_{10}$ fit uses \texttt{sklearn} (error ignorant) rather than \texttt{pyGMMis} (error inclusive). Since it doesn't account for the large errors of the pariah $i-z$, it vastly underperforms the error-cognizant $\vec c_4$ fitting. \gus{I don't think the $c_{10}$ is useful here, as it's based on a different method.} \wm{I do like quantifying gains from error inclusion, so using a GMM w/o handling errors is missing out.}
  }
  \label{fig:cf_Ncol}
\end{figure}

%> %bACC diff: [ 5.794  0.801 -1.033 -2.165]
%> sigma diff: [17.801  2.622 -3.178 -4.809]
% 2σ diff / 2: [17.512,  2.428, -2.712

% \wm{[Double-check these numbers: std instead of quantiles?]} % they are, but... I don't care a ton. 

Figure~\ref{fig:cf_Ncol} shows that for SDSS/low-$z$, single-color selection (using the optimal choices of either $g-r$ or $u-r$), selected the quenched population with $\sim 87\%$ accuracy. 
Adding a second color significantly improved accuracies (again using optimal band choices from our analysis of all possible colors), raising balanced accuracy immensely ($+5.8\%$). % ($\sim17\sigma$ significant). 
Optimal three-color selection gave a relatively small %; $2.5 \sigma$) 
increase in bACC ($+0.8\%$). 
The four-color combination of the primary color vector (no optimization of color choices) performed similarly %; $-3.4 \sigma$) 
to the best-case three-color combination ($-1.0\%$). 
The primary color vector thus serves well as a blind baseline for selecting the quenched population. 

% \gus{Suggest dropping c10 discussion. Methods that ignore errors are not interesting.}
% Using all ten colors $\vec c_{10}$ (including secondary colors---requiring use of the error-ignorant \verb`sklearn` GMM) resulted in superior selection of quenched galaxies than single-color mixtures, but worse than optimal 2--3 band selections or the primary color GM $\vec c_4$. This is largely due to the inclusion of the pariah $i-z$, which not only lands far from the 4000~\AA\ break, but also has far larger errors than the other primary colors. 
% In contrast, the primary color GM $\vec c_4$ resulted in excellent selection, despite inclusion of the pariah. 

% % % % % % % % % % % % % % % % % % 
\subsubsection{Optimal Gaussian Component Count} \label{sec:elment_count} 

Though historically galaxy classification has been binary, galaxies transitioning from RS to BC are sometimes classified as members of the green valley (GV), adding a third category. 
From an agnostic view of the color space data, components beyond two can simply be seen as an attempt to better model inherent non-Gaussianities in the populations \citep[see e.g.][]{Carretero+15}.
From an astrophysics view, galaxies quenched by different mechanisms belong to populations with distinct characteristics \citep{Peng+10,Davies+21,Dacunha+22}. 
High-mass galaxies (which are primarily mass-quenched) have different trends for mean and scatter of colors than those of low-mass galaxies (which are primarily merger- or environment-quenched) \citep{Baldry+04}, 
so modeling them with distinct Gaussians could better represent the underlying populations. 
For any of the above reasons, one may desire to model components beyond two. 
Though the Red Dragon algorithm permits any number of components $K \geq 2$, different datasets or different luminosity cuts may favor particular component counts. 

Appendix~\ref{sec:K_optimal} details our analysis of SDSS/low-$z$ for optimal component count. 
In short, though Figure~\ref{fig:bi_panel} shows clear bimodality visually, and indeed, using two components gives a fair fit to the photometric color data, using three components fits the distribution of galaxies in photometric color space significantly better. Using more than three components gave no significant improvement in fit. 
% In the BIC spirit of minimizing complexity while maximizing likelihood, we suggest using only two or three components. 
% RS definition as component count increases; with $K=2$, the quenched population may be split between components, convoluting interpretation of components. 
For simplicity of discussion and comparison, we chiefly employ the minimal two-component model in our results section, but an investigation of the effects of increasing component count is detailed in section~\ref{sec:RS_by_K}, using the Buzzard simulation.

\iffalse
\footnote{
  To quote the MICE collaboration: ``For our purposes, it is indifferent whether this green sequence corresponds to a physically distinct type of galaxies or just to the inadequacy of the Gaussian distribution to represent the red and blue populations.'' \citep{Carretero+15}
}
\fi
% Since two or four+ component models fit the data fine and have their own purposes, the Red Dragon algorithm permits any number of components $\left( \mathds{N} > 1 \right)$.  

% % % % % % % % % % % % % % % % % % 
\subsubsection{Running with magnitude} \label{sec:magnitude_run}

% \wm{[Similar intro \P\ to the last section; okay separate, I think, but consider swapping order or merging.]} 

Gaussian mixture parameters (population weight, mean color, and scatter for RS and BC) are known to depend on magnitude at a fixed redshift. 
  Nearly 100\% of bright galaxies are red, while very few of the faintest galaxies are red \citep{Baldry+04}, so component weight runs strongly with magnitude. 
  The mean color of the RS is well-known to run with magnitude \citep{Kodama+96,Gladders+98}, with the slope modeled explicitly in the RS fitting of e.g. \citet{Hao+09} and \citet{Rykoff+14}. 
  The scatter of the RS and BC also runs non-linearly with magnitude \citep{Baldry+04,Balogh+04}, though this is less often modeled. 
Therefore, {\bf a magnitude-ignorant fitting of the populations will have all parameters somewhat dependent on the limiting magnitude of the sample.} 
The brighter the sample, the higher the $f_R$, the redder the mean RS color, and the smaller the RS scatter. 
Though parameters do evolve with magnitude, how significantly does magnitude ignorance affect selection of the RS?

% The weight and mean color of RS and BC galaxies depends heavily on stellar mass and to a lesser extent local density \citep{Baldry+04,Balogh+04,Peng+10}. As luminosity correlates strongly with both stellar mass {\it and} local density, we consider first and foremost how significantly GMM parameters drift with magnitude. Using Buzzard, we quantify to what extent that drift affects component selection as a first pass on these measurements.  \gus{The above could be simplified: the mean color of RS galaxies is known to depend weakly on magnitude.  Red Dragon currently ignores this.  Is that a problem?} 

Using Buzzard, we quantify differences in RS selection between magnitude-cognizant and magnitude-ignorant models. 
Appendix~\ref{apx:quan_mag_run} shows results of this analysis. In short, {\bf while magnitude running of GMM parameters is statistically significant, their running had a relatively minimal impact on selection.} 

For thin redshift slices, selection of red sequence galaxies (where $P_{\rm RS}>.5$) was 95\% identical between the standard redshift-running and the niche magnitude-running versions of Red Dragon. 
Since this difference in selection is relatively small, we leave magnitude running out of the current version of Red Dragon in favor for prioritizing smooth redshift evolution. 
For those who wish to explicitly account for magnitude running or other secondary parameters, several workarounds exist, as detailed in section~\ref{sec:sim_sel}.

%%%%%%%%%%%%%%%%%%%%%%%%%%%%%%%%%%%%
%%%%%%%%%%%%%%%%%%%%%%%%%%%%%%%%%%%%
\section{Results} \label{sec:results}
% Confirmation: evidence and claims

Here we show results of running Red Dragon on SDSS, TNG, and Buzzard datasets. Our SDSS+TNG analysis focuses on the accuracy of selecting the quenched population whereas our Buzzard analysis highlights fit parameter evolution with redshift. 
% Short summary of main results: 
% The Red Dragon algorithm yields a high-accuracy selection of the quenched population, continuous with redshift. 
% Its multi-band selection indirectly accounts for not only sSFR, but also dust and metallicity. 

% % % % % % % % % % % % % % % % % % 
\subsection{Sloan analysis} 

% Here we discuss results from 
We run Red Dragon on Sloan and TNG data using the four primary colors derived from SDSS $ugriz$ photometry, with equation~\eqref{eqn:primary_SDSS} as the primary color vector. 
Here we present the accuracy with which Red Dragon identifies the quenched galaxy population at low (\S\ref{sec:SDSS_limits_of_hard_cuts}, including TNG) and intermediate (\S\ref{sec:SDSS_quenched_accuracy_redshift}) redshifts. 
% Section~\ref{sec:SDSS_limits_of_hard_cuts} uses the low-$z$ sample as well as TNG simulated galaxy data to compare the accuracy of selecting the quenched population. 
% using Red Dragon as compared to traditional hard cut methods (in CM and CC spaces). 
% Section~\ref{sec:SDSS_quenched_accuracy_redshift} uses the mid-$z$ sample to compare Red Dragon accuracy across the $g \rightarrow r$ redshift transition.
% , as compared to single-color selections. 
% Though quenched status alone doesn't determine membership in the RS / BC (see section~\ref{sec:beyond}), accuracy in its selection serves as a useful quantitative metric for comparing RS selectors. 

% \wm{[merge with / move down to below?]} 
Our comparison to typical CM and CC selections follow methods from the literature. 
`Typical' CM selection follows \citet{Hao+09}. After fitting the red sequence population with a Gaussian mixture in color space, we fit a line to the red sequence population (in the CM space of $g-r$ vs $m_i$), find its scatter, then select all galaxies within $2\sigma$ of the mean relation.
`Typical' CC selection follows \citet{Adhikari+20}. After finding population means via Gaussian mixtures, we draw a line between maxima (in the CC space of $g-r$ vs $r-i$), then plot a perpendicular line at the minimum likelihood point between the two components (i.e. where a galaxy is equally likely to belong to either component). 
These two methods give benchmark comparisons for standard efficiency of selection in CM and CC spaces for comparison to Red Dragon selection.

\subsubsection{Selection accuracy of the quenched population} \label{sec:SDSS_limits_of_hard_cuts}

\begin{figure}\centering
  \includegraphics [width=\linewidth] {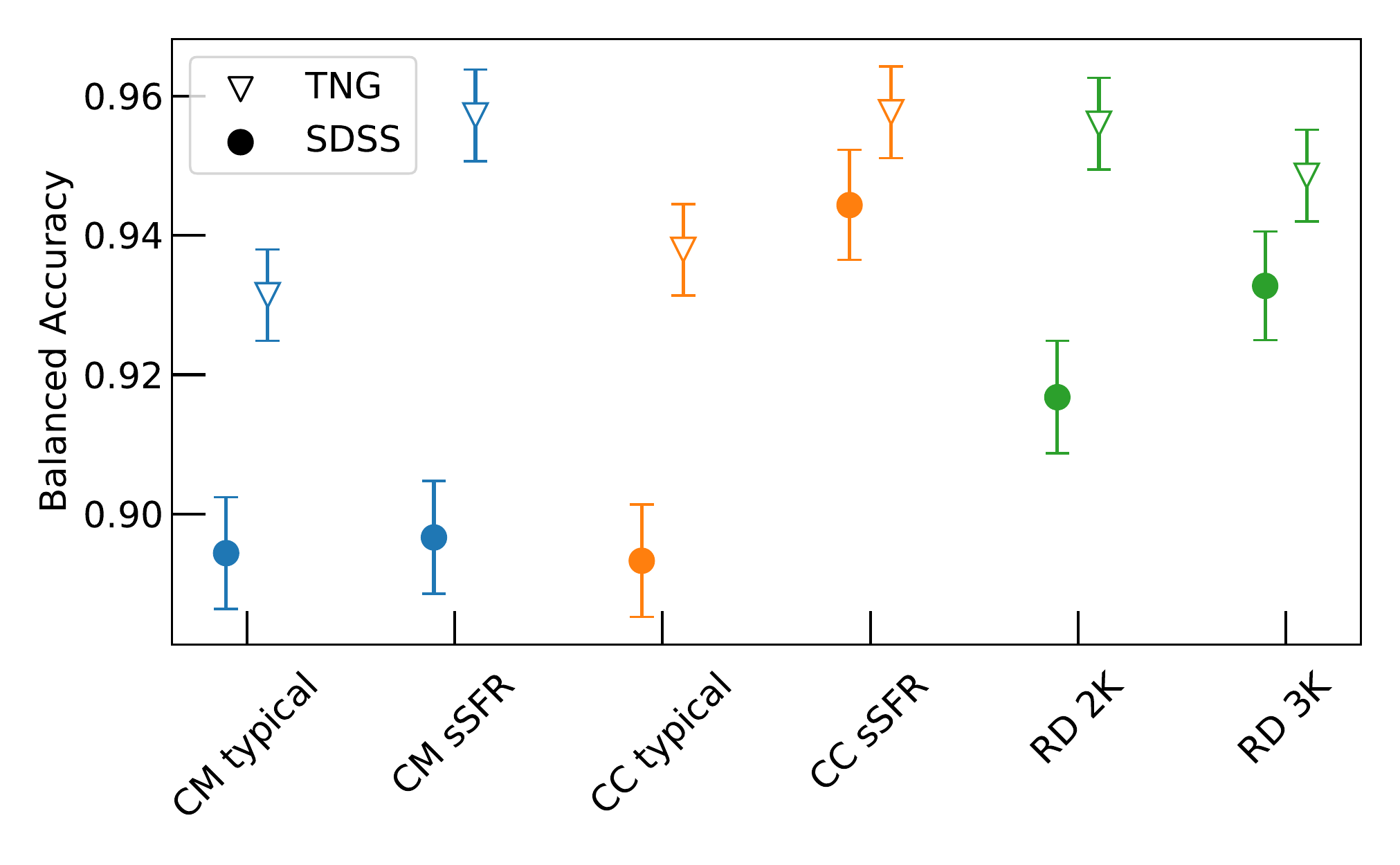}
  \caption{
    Balanced accuracy in selecting the quenched population of low-redshift ($z \doteq 0.1$) TNG and SDSS galaxies. Error bars generated from Poisson error estimates on each of the classification components. 
    ``CM'' and ``CC'' methods draw hard cuts through color-magnitude and color-color spaces respectively, whereas the ``RD 2$K$'' and ``RD 3$K$'' methods use a Gaussian mixture model with $K=2$ and $K=3$ components respectively. 
    The three-component dragons have tighter selections on the RS, yielding a lower false positive rate but a higher false negative rate, which for TNG galaxies results in a slightly decreased balanced accuracy. 
  }
  \label{fig:bACC_CM_CC_GM}
\end{figure}

To exemplify limitations of CM and CC hard-cut selections, we compare balanced accuracy in selecting the quenched population using hard cuts vs Gaussian mixtures for both the SDSS low-$z$ sample alongside the TNG sample, both at $z = 0.1$ approximately. %, as shown in Figure~\ref{fig:bACC_CM_CC_GM}. 

% \wm{[move this following \P\ above to 4.1? merge with that \P?]} 
Using methods from the literature, we make hard cut fits (labeled `typical') in CM and CC spaces. 
Next we use sSFR values to optimize fits, drawing the hard cut lines which maximize balanced accuracies (labeled `sSFR'), giving a best-case scenario for hard cut selection methods. 
Finally, we compare these fits to Gaussian mixture fitting of the populations, i.e. using a Red Dragon approach for selecting the red sequence (labeled `RD 2K/3K').

Figure~\ref{fig:bACC_CM_CC_GM} shows accuracies of these various selection types. 
For SDSS, even optimized CM selection typically incurs $\gtrsim 10\%$ error (i.e. $\sim 10\%$ of the RS and BC contain star-forming or quenched galaxies respectively) while optimized CC selection typically incurs $\gtrsim 6\%$ error, showing as did Figure~\ref{fig:cf_Ncol} that two colors (CC space) work significantly better than one (CM space). 
Hard cuts in CM and CC spaces select the quenched population more accurately in TNG than in SDSS, largely due to its more pronounced GV (see appendix~\ref{apx:SDSS_vs_TNG_hist}), with a typical error of only $\sim 5\%$ by any selection method. 

Gaussian mixtures (without any optimization from sSFR truth) perform generally on par with optimized CM and CC fits but have higher selection accuracy than CM and CC fits similarly ignorant of sSFR. 
Given a spectroscopic sample of galaxies, where sSFR values are known, Figure~\ref{fig:bACC_CM_CC_GM} shows that one could define hard cut selections of the RS which would have accuracies similar or superior to a GM selection of the RS. However, at redshifts where the RS \& BC are not well defined from spectroscopy, or at any redshift where sSFR values are unknown, a GM would give superior selection of the quenched population. 

% \gus{To me the basic points of this section's analysis are: i) RD is able to efficiently identify quenched galaxies in populations with rather different color-space structure (TNG vs. SDSS); ii) RD works as well as color cuts at low-z to identify quenched galaxies, and iii) the quenched accuracy of RD is continuous with redshift.} 

\subsubsection{Redshift continuity of quenched galaxy selection} \label{sec:SDSS_quenched_accuracy_redshift}

Here we investigate Red Dragon's accuracy in selecting the quenched population as a function of redshift. The SDSS mid-$z$ sample centers around the transition redshift of $z \sim 0.4$, where the 4000~\AA\ break moves from $g$~band into $r$~band. 
% We compare Red Dragon accuracies to (redshift evolving) single-color selections. 

% To what degree quenched status can be determined by hard cuts, as c.f. Red Dragon

\begin{figure}\centering
  \includegraphics[width=\linewidth]{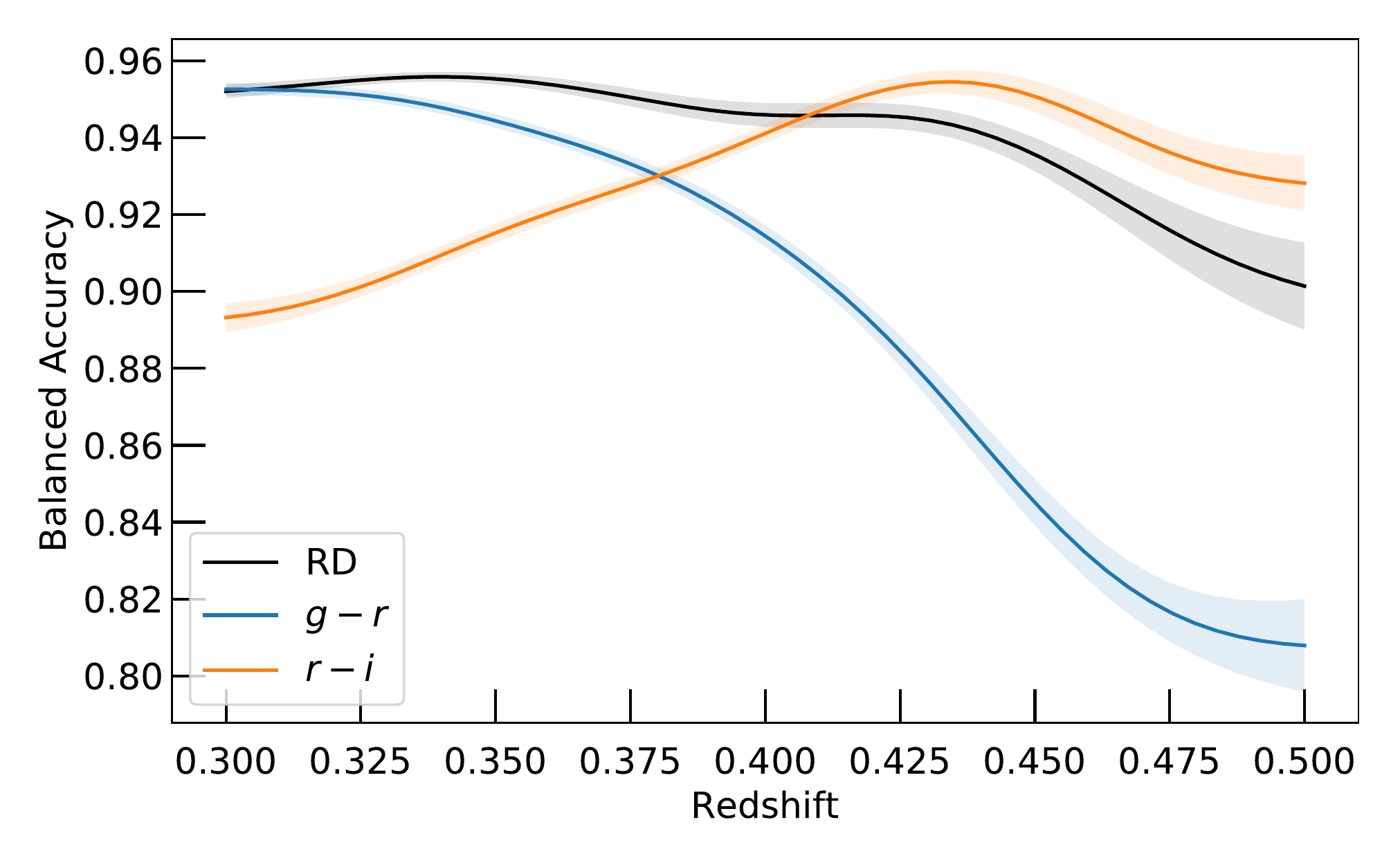}
  \caption{
    Balanced accuracy in selecting the quenched population of bright SDSS galaxies: 
    Red Dragon (RD; black) performs similarly or superior to hard cuts in single colors. 
    Bootstrap $\pm 1 \sigma$ error shown with transparencies; these increase with redshift chiefly due to decreasing number counts (rather than from increased intrinsic scatter). 
    Values localized by Gaussian kernel (width $\sigma_z = .02$). 
    % Within $3\sigma$, RD performs equivalent to or superior to single band selection. At the transition redshift ($z \doteq .38$, where $g-r$ dies out in favor of $r-i$), RD performed $>6\sigma$ superior to either of the lone colors. 
    % see section 6.2 of R14: ``With SDSS data, we use g − r for z < 0.35 and r − i for z > 0.35.'' 
  }
  \label{fig:SDSS_Z}
\end{figure}

% Hard cuts in single (or even multi-) color spaces yield limited accuracy, especially over large spans of redshift.
% introduce and unpack figure
Figure~\ref{fig:SDSS_Z} compares accuracy in selecting the quenched population between Red Dragon and two (redshift-evolving) choices of single-color cuts for defining the RS. 
As the 4000~\AA\ break passes from $g$~band to $r$~band near $z=0.36$ (thus affecting the values of $g-r$ and $r-i$), the ability of $g-r$ to select the quenched population wanes while that of $r-i$ waxes, as expected. 
If using single-color selection, $z=0.38$ would then be the best redshift to transition from selecting the RS with $g-r$ to selecting with $r-i$ (if your goal is to select the quenched population with greatest fidelity). 

In comparison to these single-color selection methods, Red Dragon performs similarly to best-case single-band selection (within $3\sigma$), with vastly superior ($>6\sigma$) accuracy across $z=0.38$, the optimized transition redshift. We note here that the high-redshift side of the plot has significantly lower number counts, and lacks statistical power compared to the low-redshift side. 
If using single-color selection of the RS, optimal transition redshifts between colors are somewhat subjective. 
The initial calibration of the RS by RedMaPPer transitions from using $g-r$ to using $r-i$ at redshift $z=0.35$ (the 4000~\AA\ transition redshift; also the point below which the survey is volume limited and above which is magnitude limited). 
However, they found that the redshift for which single-color richnesses $\lambda_{g-r}$ and $\lambda_{r-i}$ equaled their multi-color richness $\lambda$ was at redshift $z \doteq 0.42$ (see their Figure~28), so a reliable $\lambda_{\rm col}$ definition would use $z \doteq 0.42$ as a transition redshift. 
Neither $z=0.35$ nor $z=0.42$ match the single-color transition point highlighted by Figure~\ref{fig:SDSS_Z} of $z=0.38$ (the redshift at which trading off from $g-r$ to $r-i$ maintains the highest quenched population selection accuracy). 
Since these redshifts each have sound reasoning for their use in defining a single-color RS selection, no universal transition redshift stands out. 
This leaves single-color RS selection transitions as messy at best, favoring the objectivity of multi-color analyses. 

% Though accurate sSFR information for selecting the quenched population doesn't exist for later redshift transitions, we expect similar behavior with the r to i transition and so forth. 

% Even with perfect knowledge of the quenched population a hard cut in photometric space still evokes some 5--15\% error (see Figure~\ref{fig:bACC}).

Red Dragon preserves accuracy in selecting the RS across redshift transitions while maintaining a continuous red fraction (by construction). 
This then evades the discontinuities inherent in swapping bands,% (see section~\ref{sec:z_drift}), 
continuously selecting RS galaxies with high fidelity.

% % % % % % % % % % % % % % % % % % 
\subsection{Buzzard Flock analysis} \label{sec:Buzzard} 

Extending our analysis to a wider redshift range, we turn to the synthetic galaxy catalogs of the Buzzard Flock with the three primary colors of equation~\eqref{eqn:primarycolor}. 
After highlighting how fits interpolate across redshift (\S\ref{sec:fit_interp}), we discuss how the RS definition varies as component count increases from two to four (\S\ref{sec:RS_by_K}). % , compare to RedMagic-selected galaxies, and finally discuss its possible utility in creating a better mass proxy. 

\subsubsection*{} % consider adding section here like "introduction to the buzzard universe"? 

The Buzzard universe is a statistical replica of a deep-wide galaxy survey built from galaxy color distributions measured as a function of local cosmic overdensity \citep{Hogg_SDSS_2004}.
The ADDGALS method is trained empirically at low redshifts and extrapolated to high redshifts using a spectral energy distribution template approach \citep[for details, see][]{Wechsler+21}.  While the method reproduces well the counts and two-point clustering statistics of galaxies \citep{DeRose+21}, behaviors of the Buzzard universe at high redshifts are less rooted in observation than those at low redshift.

With a sample size of 94M galaxies (see table~\ref{tab:datasets}) we are able to extract precise estimates of all model parameters. However, the statistical errors shown below are lower limits, in that systematic variations caused by a different galaxy catalog construction method (see e.g. MICE \citep{Carretero+15}, cosmoDC2 \citep[populated by GalSampler algorithm][]{Hearin+20}, etc.) remain to be investigated. 
% \gus{Forward reference your DES paper here?} 

% \wm{[Could move this \P\ forward to mean color sub-section, but I like having it here, since it applies to multiple of the sections to follow.]} 
For RS mean and scatter, we compare to \citet{Hao+09} (SDSS catalogue) and \citet{Rykoff+14} (DES catalogue). 
\citet{Hao+09} fit SDSS data using an error-corrected GMM in $g-r$. With that selection for blue and red, mean colors as a function of $i$ band were measured along with scatters, all as functions of redshift. 
\citet{Rykoff+14} fit data using a multivariate error-corrected GM in the primary colors of $griz$ (see equation~\ref{eqn:primarycolor}). Their algorithm iteratively selects the RS, measuring a slope of its color as a function of $z$~band, giving a redshift-continuous fitting across redshift much like Red Dragon. 
This method was then applied to DES~Y3 data to provide an observed RS fit (E.~Rykoff, private comm.). 
% NOTE: I've requested Eli upload this to Zenodo so I have a dataset to point towards, but until then, I plan on still using R14 as the paper to point towards, since it describes the method of fitting, even if not the actual output parameters used in the below fitting. 
Both methods only fit the RS, so no information on weight nor any fits for the BC are available for comparison to Red Dragon.

\subsubsection{Redshift Evolution of Two-Color GMM Components} \label{sec:fit_interp}

This section details a fitting of the evolving GMM parameters across redshift for Buzzard photometry. 
% After fitting each redshift bin, Red Dragon interpolates between (eschewing outliers). This fitting then characterizes the RS in a continuous manner, resulting in a red fraction with no redshift discontinuities. 
% 
The galaxy sample is magnitude-limited using the  redshift-evolving cut of $0.2 \, L_{*,i}(z)$ from \citet{Rykoff+14} within the redshift range $0.05 < z < 0.84$.  The galaxies are divided into narrow cosmological redshift bins of width $0.025$, resulting in counts per redshift bin of 60k to 7.5M galaxies.  
% The following fits used 50 bootstrap realizations of Red Dragon parameter fits to measure consistency of GMM parameters $\vec \theta$ within each redshift bin from $z=0.05$ to $z=0.84$ (bin width $dz \sim 0.025$). 
Red Dragon is run on 50 bootstrapped samples of size $10^4$ (undersampling for the sake of speed, efficiency, and easing computational burden); the resulting median parameters $\theta(z_i)$ and $\pm1\sigma$ quantile range for each bin are shown in the figures below. 
% \meh{[I didn't use $>10^4$ b/c it's already time consuming, and doing 10 to 100 times more galaxies would consume much more energy and time, and pyGMMis doesn't handle well being given too many datapoints at once. I sacrificed precision of fit for the sake of having faster and more efficient resampling.]} 
% Why did I undersample? 
% 1) pyGMMis doesn't handle well being given too many points. 
% 2) would take more CPU hours to run, which means more time and energy spent on this, when a quick pass is enough to get the feel for it. 
The discrete redshift parameters are then interpolated using KLLR with a Guassian kernel of width $\sigma_z = 0.02$, shown as lines in the figures below.

% Since GMM parameters run with stellar mass (see \S\ref{sec:optimal_bands} and \S\ref{sec:elment_count}), the magnitude-ignorant fitting of parameters we find depends on the luminosity cut of the sample. We use the redshift-evolving magnitude cut from \citet{Rykoff+14} of $0.2 \, L_*$ to limit our analysis to bright galaxies. 
% This means that our parameterizations of $\theta(z)$ are somewhat like weighted averages, giving the roughly the behavior of a median luminosity galaxy rather than general behavior for any luminosity. 

\iffalse
\begin{table}\centering
  \begin{tabular}{r|l}
  bin width & $dz \sim .025$ \\
  bin extent & $z|[.05,.84]$ \\
  $N_{\rm bootstrap}$ realizations & $50$ \\
  luminosity cut & R14 (see \S2.1) \\
  $\min(N_{\rm gal})$ & $59 661$ \\
  $\max(N_{\rm gal})$ & $7 554 935$ 
  \end{tabular}
  \caption{
    \wm{could have as table something like this; not sure how important each element is}
    information regarding the bootstrapping for Buzzard fits
  }
  \label{tab:bootstrap}
\end{table}
\fi

\begin{figure}\centering
  \includegraphics [width=\linewidth] {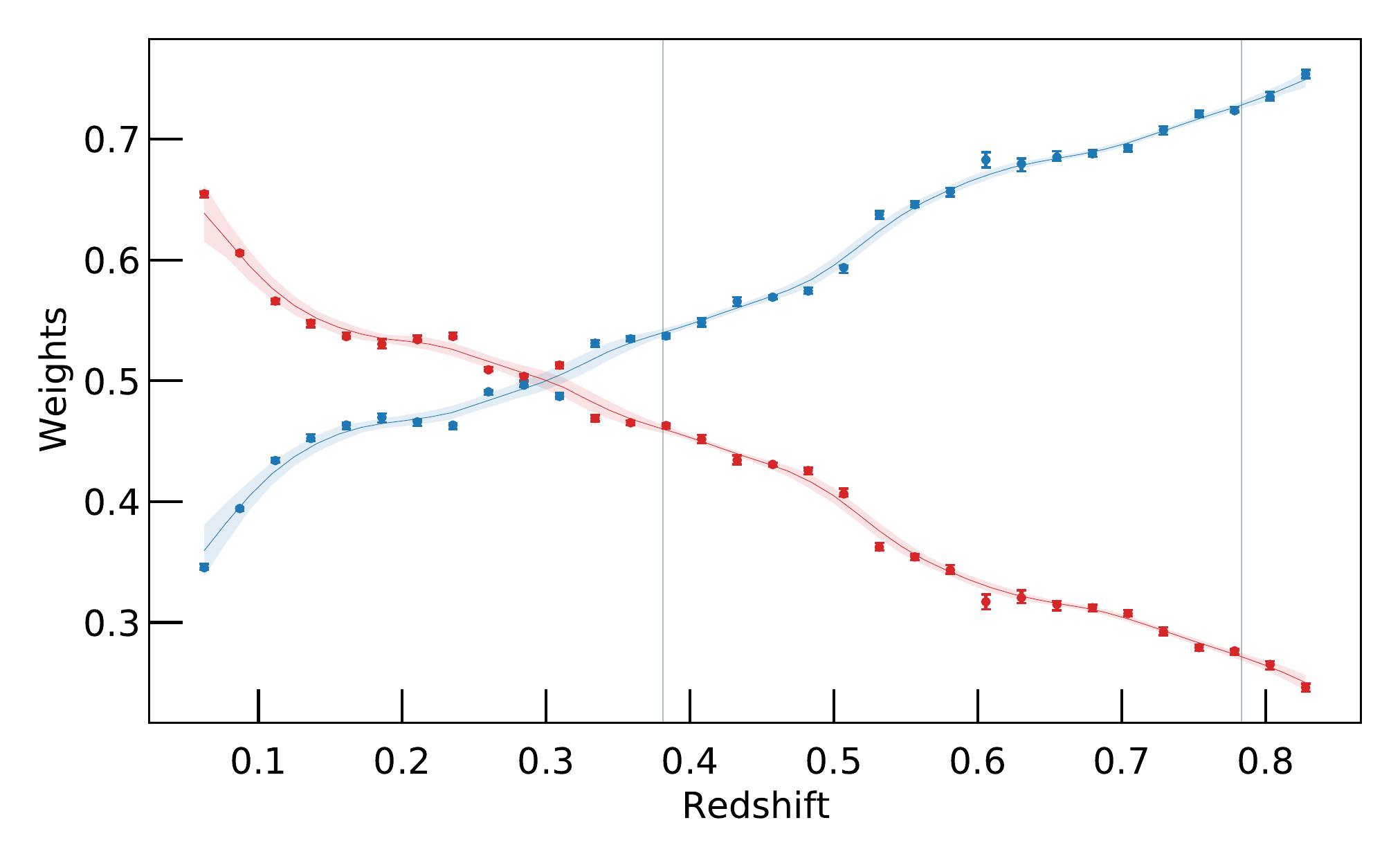} % {figs/bootstrap_w} % {figs/Bz_R14_2K_w.pdf}
  \caption{
    Component weight for RS (red) and BC (blue) as a function of redshift for the Buzzard flock. % (magnitude limited to $L > 0.2 L_*(z)$). 
    Points show parameter fits from individual redshift bins with bootstrap errors shown as error bars. 
    Fit line interpolation smoothed with KLLR (Gaussian kernel width $\sigma_z = 0.02$) with uncertainty in fit shown as transparencies about the line. 
    Vertical grey lines indicate transition redshifts of the 4k\AA\ break from 
    Table~\ref{tab:4kA_break}. 
  }
  \label{fig:Bz_w}
\end{figure}

% % % % % % % % % % % % % % % % % % % % % % % % 
\paragraph*{Component Weights.}

A variety of deep observations of the real universe indicate that star formation rates per unit baryon mass were much higher in the past \citep{Madau_1996, Connolly_1997, Madau_Dickinson_2014}. % and have been declining over the last 10~Gyr
% https://academic.oup.com/mnras/article/413/1/80/1059944 ? not sure which papers you're talking about here... 
Astrophysically speaking, while the Butcher-Oemler effect of reddening over time \citep{Butcher_Oemler_1978} applies primarily to galaxy clusters, the entire population of galaxies ages and tends to redden as a whole. 
Quenching is nearly a one-way process for galaxies \citep[many models ignore the reverse direction entirely, e.g.][]{delaBella+21}, implying that the only way to decrease red fraction over time is to create new blue galaxies. 
Since the peak of cosmic noon was at $z \sim 1.9$ \citep{Madau_Dickinson_2014}, galaxy samples below this redshift should redden over time. 
We therefore expect Red Dragon, when applied to Buzzard, to extract a RS weight that declines with increasing redshift (i.e. the population becomes bluer with increasing redshift). 
% "bluen" is a real word, albeit obscure 

In good agreement with this expectation, Figure~\ref{fig:Bz_w} shows that the RS weight consistently decreases with redshift, ranging from roughly 70\% at redshift $z=0$ down to 25\% at the highest redshift of $0.84$. 
Since the red fraction is highly luminosity dependent (as discussed in appendix~\ref{apx:quan_mag_run}), % ; see equation~\ref{eqn:fR_mag_run} 
one should rememeber that the weight reported here represents a weighted average of all galaxies above $0.2 L_\ast(z)$, which will be dominated by magnitudes near the cutoff.  Choosing a brighter magnitude cutoff would uniformly raise the RS weights, and vice-versa.

\begin{figure}\centering
  \includegraphics [width=\linewidth] {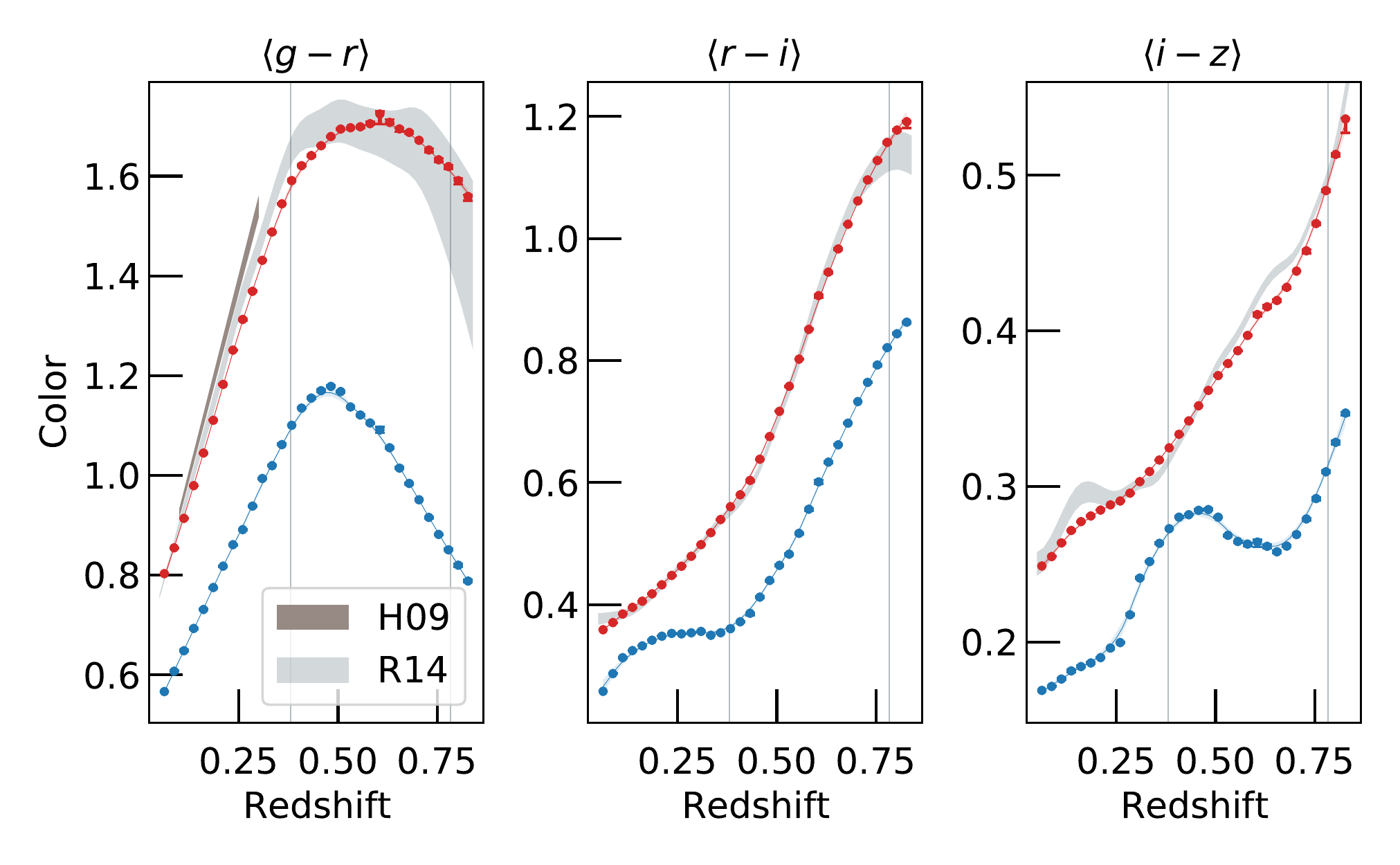} % {figs/bootstrap_mu} % {figs/Bz_R14_2K_mu.pdf}
  \caption{
    Points show mean colors for each GMM component (RS \& BC) for Buzzard galaxies, using the same coloring as Figure~\ref{fig:Bz_w}. RS mean  measurements from \citet{Hao+09} (SDSS, $g-r$ only) and \citet{Rykoff+14} (DES~Y3, all colors) are shown for comparison. 
    Due to their dependence on magnitude, we present ranges from $0.2 \, L_*$ (the magnitude limit of Buzzard; lower edge of transparencies) up to $L_*$ (upper edge of transparencies).  respectively. 
    % \gus{suggest y-axis label: mean color, $\bar{\mu}_\alpha$.} 
    % \wm{I'm not a fan of these proposed axis label changes, but could be persuaded. I like how simple and concise the y-axes labels are currently, with the more technical definitions on top.}
  }
  \label{fig:Bz_mu}
\end{figure}

The bootstrap uncertainties are typically quite small, but there is an increase near $z = 0.6$ (seen somewhat if figure~\ref{fig:Bz_mu} and especially in the correlations of \ref{fig:scat_corr}). Here, rather than giving slight variations around a single fit as at earlier redshifts, pyGMMis at this redshift debates between two distinct fits: one with wider scatter (and low correlation) and one with narrower scatter (and high correlation), each of which having differing RS weight. These two modes exist in very few of the previous bootstrap realizations, but near $z = 0.6$ make up roughly half of the fits. Since Bootstrap resampling yields two discrete modes, the overall uncertainty is relatively large compared to single-mode redshift bin fits. 

% \wm{[I'm not certain as to {\bf why} pyGMMis suddenly is confused between these two modes, but I think it may be because there's less of a separation between RS and BC / the GV is more populated at $z=.6$ which would make the RS more likely to be more elongated / narrower, thus favoring that high-correlation mode. But I'm not too certain about why, so I don't think it's worth making a statement on it.]}
% \wm{[This {\it does} correspond to the last time $\sigma_{g-r} < {\rm median}(\epsilon_{g-r})$, or in other words, the last time that for any color's scatter was larger than the typical errors of that color. So that may have something to do with it. Maybe once all colors became error-dominated, something shifted in how the code operates? But ellipses beyond that redshift are sometimes clean, looking like the $z < .6$ ellipses, so I don't have high confidence in that explanation. But it could be related to why / causally explain why the scatter takes off in each of the colors at that same redshift.]}

% This is largely caused by a peculiarity in the distribution of galaxies near this redshift: the blue cloud has significantly more galaxies towards its bluer end, in a nearly separate lobe, tempting the fit towards a more stretched out orientation than its neighboring redshift bins. Resampling around this redshift then yields a strong variety in BC fit parameters, indirectly affecting RS fit parameters as well. % Not so sure about this... 

% % % % % % % % % % % % % % % % % % % % % % % % 
\paragraph*{Mean Colors.}

Figure~\ref{fig:Bz_mu} shows the redshift evolution of the mean colors of the two components. The three panels each show measured BC and RS means in different colors with comparisons to observations in shades of grey (\citet{Hao+09} only fit $g-r$ at low redshift). Since these observations were magnitude dependent (whereas the Red Dragon fitting of Buzzard is magnitude independent), mean colors between $L_*$ (upper bound of transparency) and at $0.2 \, L_*$ (limiting luminosity of Buzzard galaxies; lower bound of transparency) are shown for comparison. 
Because over two-thirds of Buzzard galaxies fall between these two limits, we expect our mean color fitting to also be between these bounds as well. However, the Buzzard photometry differs from photometry input for the comparison observations, so the fits will differ somewhat. 
Note that each color has different vertical scaling: $\langle g-r \rangle$ spans the largest range while $\langle i-z \rangle$ spans the smallest range, exaggerating its features.

At transition redshifts (see Table~\ref{tab:4kA_break}), the slope of mean color with respect to redshift changes rapidly, % \gus{The gradients in mean color generally are not maximized at the filter transitions, so what do you mean by ``color changes rapidly''?  The mean behaviors all look pretty smooth, as they should be since they're differences in integrated spectral magnitudes.}, 
necessitating narrow redshift analysis bins and careful fitting. 
% Away from these transition redshifts, mean color drifts near linearly, so long as one of the bands composing the color contains the 4000~\AA\ break. 
% Away from these transition redshifts, mean RS colors drift near linearly with redshift, especially as the 4000~\AA\ break glides through one of the magnitudes composing a color. 
Mean colors for both RS \& BC follow a general shape of rising as the 4000~\AA\ break enters the color's minuend, then falling as the break enters the color's subtrahend, as expected. 
% Differences in filter shapes (e.g. the $r$-band transmission curve is more symmetric than that of $g$-band) cause differences in the duration of peak redness and the shape of colors at transition redshifts. % Wm: Save this discussion for the DES paper! 

% \gus{What's driving the (relatively) complex behavior of the BC in $i-z$?} \wm{no idea ... can look into stellar / galactic spectra there.} % Save this for the DES paper! 

Compared to observed mean colors, Buzzard shows a bluer red sequence for $z \lesssim 0.45$ in $\langle g-r \rangle$, albeit by a small margin. Comparing to the \citet{Rykoff+14} method, deviations from observations are consistently $<0.034~{\rm mag}$ (only $0.005~{\rm mag}$ deviant on average). 
% when considering the spread between this sample's faintest galaxies (dashed lines show $0.2 \, L_*$) and $L_*$ galaxies (solid lines), 
For $z \gtrsim 0.45$, colors vary more with magnitude (steeper RS slope in CM space), such that Buzzard mean colors land between the $0.2 \, L_*$ and $L_*$ observations. 
% \gus{Expand this comparison to previous work and speculate on source of the small mean differences. You only have two paragraphs to explain mean color behavior, which seems a bit thin given there are two galaxy components and three bands. What about linking the z-behaviors of the different colors?}

If the mean spectra of RS and BC galaxies had no time evolution, then the same general shape of each curve in figure~\ref{fig:Bz_mu} should appear for each color, shifted by roughly $\Delta z = .4$ (and vertically scaled somewhat for differences between bands). 
We see this to some extent, e.g. with the rising of the RS $\langle g-r \rangle$ in $z|[0,0.4]$ mirroring the rising of the RS $\langle r-i \rangle$ in $z|[.4,.76]$. 
Similarly, the BC bump in $\langle i-z \rangle$ at $z \sim 0.5$ mirrors the $z \sim 0.1$ BC bump in $\langle r-i \rangle$ (the vertical scaling difference between the axes belies their similarity). 
As our analyses extended to higher redshifts in future work, we will find similar universal features or find time variation in RS and BC populations.

% % % % % % % % % % % % % % % % % % % % % % % % 
\paragraph*{Color Covariance.}

\iffalse % covariance matrix unintuitive to read. 
\begin{figure}\centering
  \includegraphics [width=\linewidth] {figs/Bz_R14_2K_Sigma.pdf}
  \caption{
    Redshift evolution of lower triangle of covariance matrix between colors (labels show which colors correspond to columns and rows; vertical direction indicates magnitude of covariance element). 
    Coloring of components and interpolation as detailed in Figure~\ref{fig:Bz_w}. 
  }
  \label{fig:Bz_Sigma}
\end{figure}
\fi

\begin{figure}\centering
  \includegraphics [width=\linewidth] {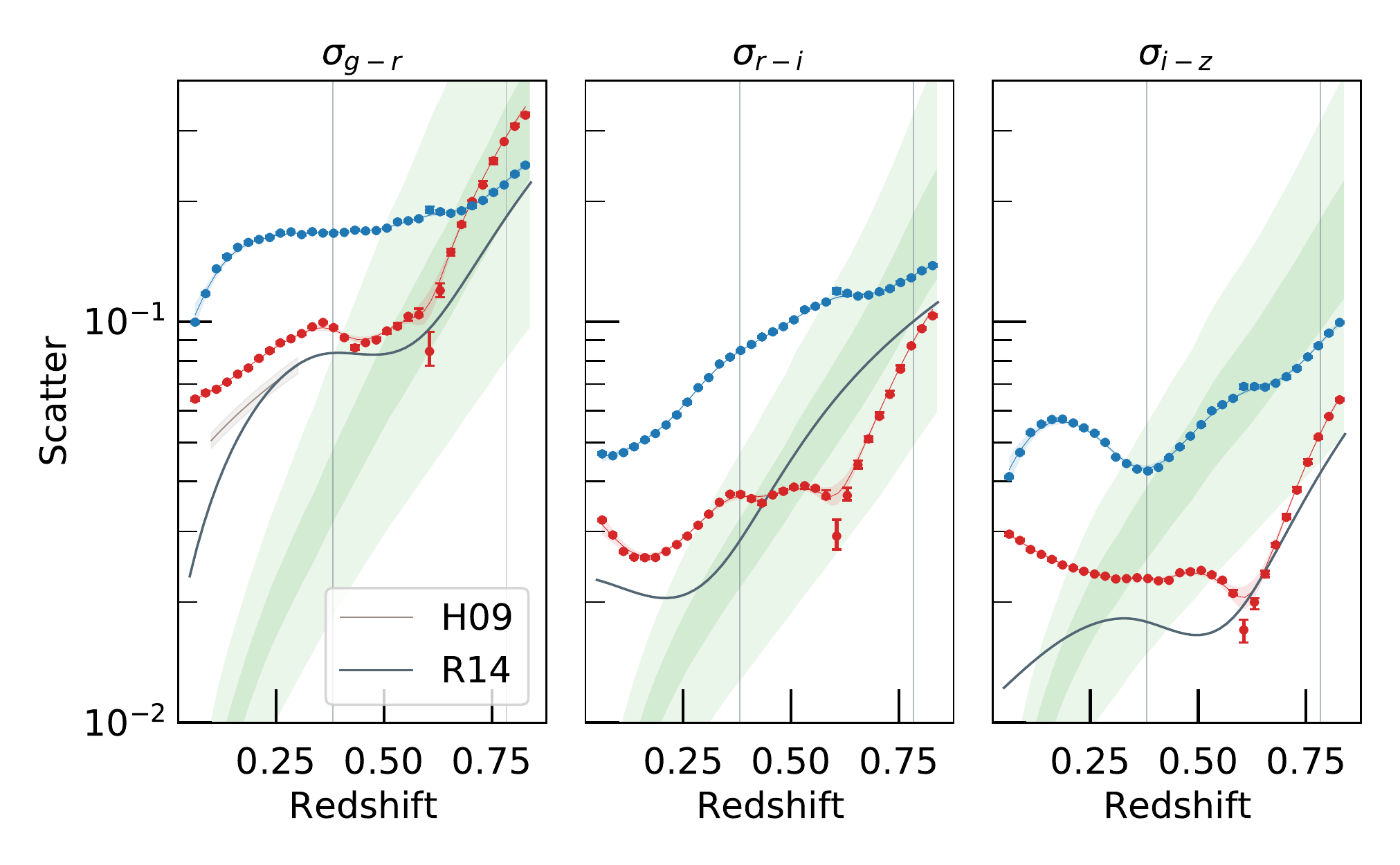} % {figs/Bz_R14_2K_scat_boot} % {figs/bootstrap_scatt} % {figs/Bz_R14_2K_scat_new}
  \includegraphics [width=\linewidth] {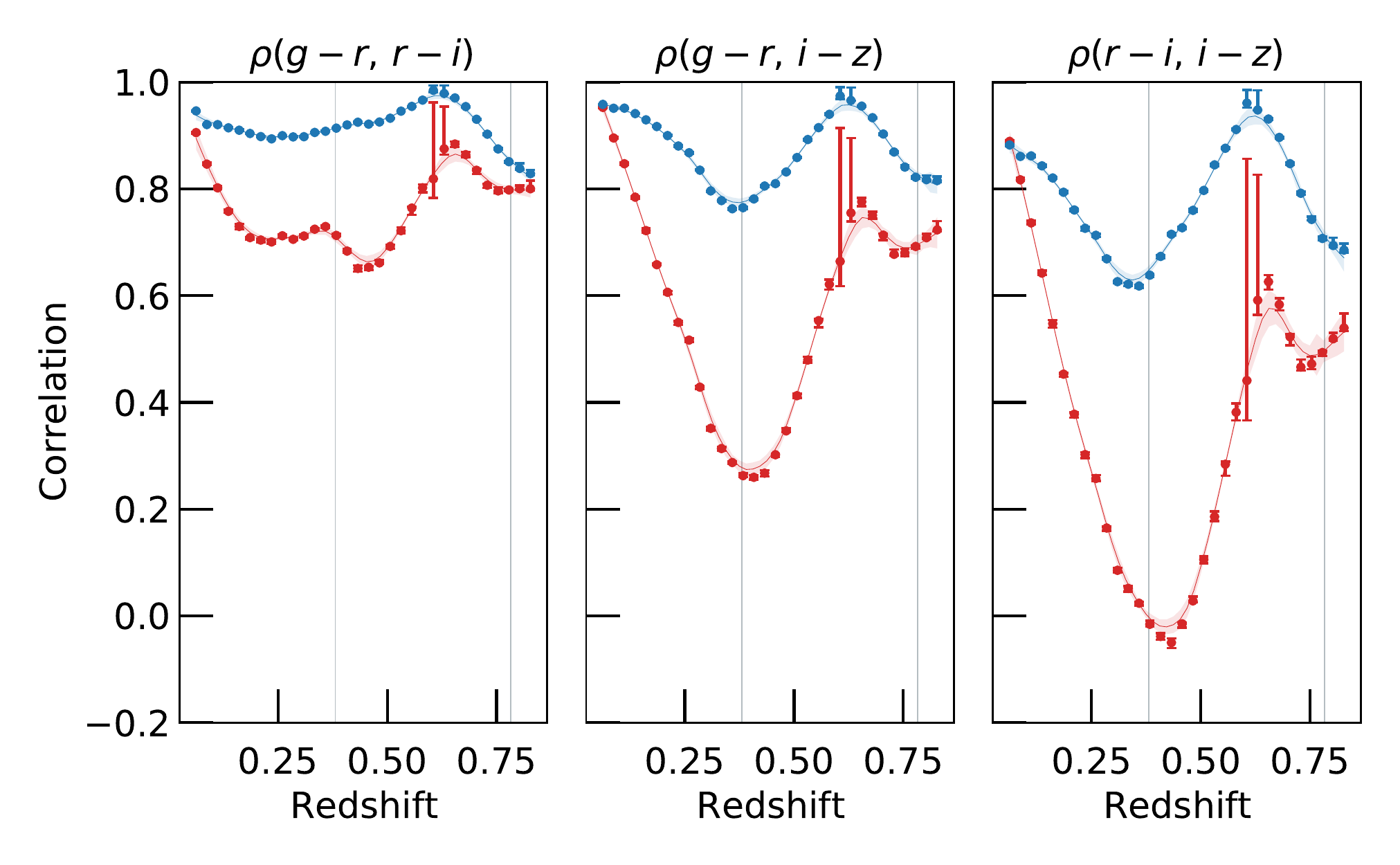} % {figs/bootstrap_corr} % {figs/Bz_R14_2K_corr_new}
  \caption{
    Intrinsic color scatter (upper panel) and correlations between colors (lower panel). 
    % \wm{perhaps add note: ``Buzzard shows drastically smaller correlations between colors than expected from observations''}
    % Comparisons to RS scatters from \citet{Hao+09} and \citet{Rykoff+14} shown as light and dark red lines respectively. 
    Green transparencies give one and two-sigma quantile distributions of photometric color errors in Buzzard. 
    Coloring of components and interpolation as detailed in Figure~\ref{fig:Bz_w}; comparison to observations as in Figure~\ref{fig:Bz_mu}.
    %
    % \gus{Suggest using y-axis label, ``Intrinsic Correlation, $\sigma_{\alpha,\beta}/(\sigma_\alpha \sigma_\beta)$ '' for bottom, then you can remove $\rho$ from the top labels.  Also ``Intrinsic Scatter, $\sigma_\alpha$'' for top, then you can use the same top labels as Fig 7. } 
    % \wm{Since correlation would have different labelling style anyway, I'm not eager to change things just so $\mu$ and $\sigma$ have the same top labels.}
    % Since H09 uses only g-r, no correlations between colors are available here. 
    % Note that there are no dashed lines here, since R14 didn't allow scatter nor correlation to drift with magnitude. 
  }
  \label{fig:scat_corr}
\end{figure}

The top row of Figure~\ref{fig:scat_corr} shows the redshift evolution of the intrinsic scatter in each color for both GMM components. Showed also are the fits from observations in grey (as before but without magnitude dependence) as well as $\pm1\sigma$ and $\pm2\sigma$ median quantiles for color errors, shown as green transparencies. 
The bottom row of Figure~\ref{fig:scat_corr} shows intrinsic correlations between colors inferred from the covariance matrix, giving $\rho(g-r, \, r-i)$, $\rho(g-r, \, i-z)$, and $\rho(r-i, \, i-z)$ from left to right. 

% While the 4000~\AA\ break drifts through a color's minuend, we see a vast increase in scatter, by a factor of two to five. \gus{Vast?  The rise can be large, yes, but it's also fairly gradual.  And this is RS only, right?  Best to describe RS and BC behaviors separately.} 
In accordance with expectations, at redshifts where the color's minuend contains the 4000~\AA\ break, the RS has significantly lower scatter than the BC (by a factor of $\sim 2$), indicating a tighter population. 
% In particular, RS $g-r$ scatter increases drastically as redshift nears the point of the 4~k\AA\ break leaving $r$~band (with its upward trajectory beginning near the point when the 4~k\AA\ break leaves $g$~band). In general (though not globally so), scatters increase considerably with redshift. 
Due to an increase of photometric errors at higher redshifts (as shown by the green transparencies), it's difficult to accurately constrain population scatters. Beyond redshift $z \sim 0.4$, the median photometric error exceeds the intrinsic RS scatter, making it difficult to measure. 

Compared to observed RS scatters, Buzzard scatters are generally wider by a factor of 1.5 (consistently within a factor of three above or below). 
Running of the RS mean color with magnitude means that a magnitude-ignorant fitting of the population scatter would find larger values than a magnitude-cognizant model, so Buzzard's generally wider scatters are expected. 
% Since Red Dragon ignores luminosity, this is expected, since the running of RS with magnitude would increase the total width of the RS. However, the strength of this magnitude dependence increases with redshift, 
% (the RS slope is negligible compared to RS scatter at these low redshifts). 
% so the low-redshift discrepancy is less expected than the high-redshift discrepancies.  \gus{Can you be a bit more quantitative here?  Also DES-Y3 scatter in $r-i$ is larger than that of Buzzard at high z.} 
% \wm{nix? [fit failure discussion] The increased scatter at lower redshifts also reflects a failure of fitting: though the RS and BC ellipses lie essentially in the correct places [check if true], their weights are off, such that the reddest portion of the BC is incorporated into the RS component. ... [I'm not sure on this. the fit seems relatively good in CM spaces, and doesn't really look horrible in CC spaces either... I should be cautious here, not putting down the fit too much, especially when I'm not certain it's actually a horrible fit.]}
%\gus{May be good to remind the reader that a 2-component color-space GMM is not guaranteed to identify the same RS as, say, redmagic.} 
Since \citet{Rykoff+14} only fit the RS, a Red Dragon fitting of the RS (which also accounts for and fits the BC) is not guaranteed to identify the same exact RS. This will be further investigated in our papers to come, analyzing DES data. % \gus{Comment that you will follow up on this in a forthcoming DES paper?} 

\paragraph*{} % for spacing purposes, to transition from scatters to correlations. Could have them as actual section headers instead of having both under "covariance", but they're part of the same matrix, so I prefer keeping them together. 

While measures of the RS scatter in various colors are available in the literature \citep{Hao+09,Rykoff+14}, to our knowledge there is no published work on measuring the intrinsic scatter of the BC nor covariance among primary (or otherwise) colors for RS and BC galaxies. 
If this is correct, the BC scatter and full color covariance as a functions of redshift constitute empirically unexplored territory. 
Our Buzzard measurements are then establishing a first estimate of these quantities, albeit one more likely to reflect that of the true galaxy population at low redshift than at high redshift.

Though intrinsic color correlations within each of the RS and BC are expected to be $\gtrsim 90\%$ \citep[][]{Rykoff+14}, % mentioned in text vaguely, but I could cite Eli / DES data / back of the envelope theory with BC03 models 
the GMM fitting of Buzzard usually has lower (and occasionally even has negative) correlations between colors. % , yet another peculiarity of the Buzzard flock. This may be a result of Buzzard's being tuned for low-redshift SDSS galaxies---noise from low redshifts gets copied to higher redshifts, and these low correlations could be one such case. % Eli: ``I think the problem with Buzzard is that it’s tuned from low-z SDSS data, and so noise there just gets copied everywhere.'' (re these plots) 
Since the simulated photometric errors in Buzzard were of similar order to the RS scatter for $z \gtrsim 0.4$ (see the green transparencies of Figure~\ref{fig:scat_corr}'s upper panel), these correlations represent more so relations between photometry than intrinsic correlations within the RS. In comparison, the BC has a width larger than the photometric error at all but the highest redshifts, and accordingly, we see high correlations, around 80\% to 90\% across redshift. 
% While redMaPPer usually freezes correlations at 90\%, this run let correlations run free, and aren't to be trusted as intrinsic correlations of the RS population. Similar to Buzzard, the correlations largely measure photometry correlations instead.  \gus{What about adding the color errors expected from assigned magnitude errors (for an 0.2 Lstar galaxy) at $z=0.25, 0.5, 0.75$ to the top panel of Fig 8?  These could be used to frame the discussion about how hard it is to extract intrinsic covariance.  }  

These fits show that Red Dragon can map out the RS and BC to high redshift (spanning across multiple transition redshifts), 
% (of the 4k~\AA\ break; see table~\ref{tab:4kA_break}), 
continuously parameterizing important aspects of each population. 
As future studies create more complete samples of galaxies to higher redshifts, Red Dragon will be able to detail population characteristics smoothly across even wider redshift ranges.

\subsubsection{RS robustness to component count} \label{sec:RS_by_K}

\begin{figure*}\centering
  \includegraphics [width=\linewidth] {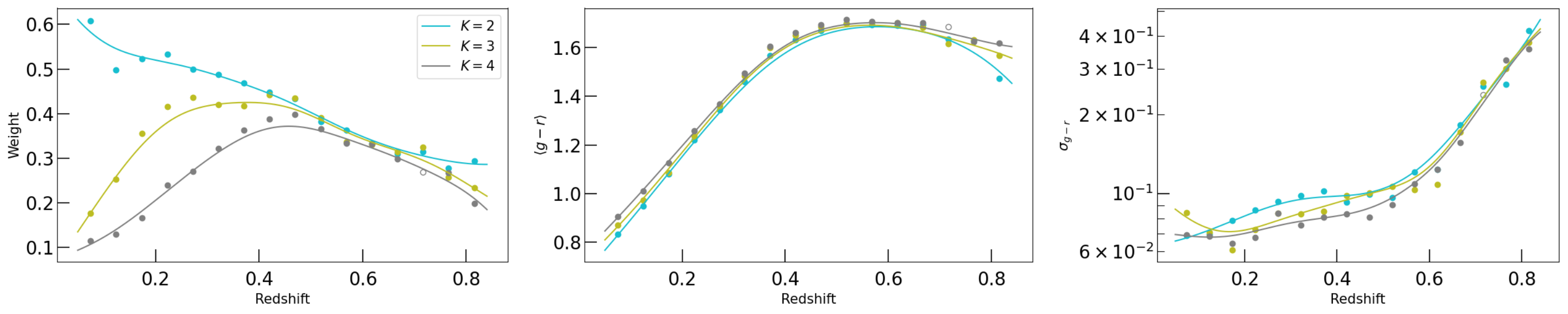}
  \caption{
    RS fit variation between component count K, shown for photometric color ($g-r$). % These sub-divisions show good consistency within the RS, indicating the sub-division of the BC moreso than the RS with the addition of new components. 
    {\bf Left}: Red fraction decreases significantly with the addition of more components below $z=0.4$, but for higher redshifts, weight varies less than $5\%$ (typically by $\sim 1\%$). 
    {\bf Middle}: Mean RS color varies by less than $0.05~{\rm mag}$ on adding a component (typically by $\lesssim 0.01~{\rm mag}$). 
    {\bf Right}: Scatter in the RS decreases with the sub-division of additional components, on average by $<4\%$, consistent to less than a factor of $1.5$ across all redshifts. 
  }
  \label{fig:Bz_Kcf}
\end{figure*}

Here we investigate how Red Dragon identifies the RS in models with more than two components. 
On adding additional components to a Gaussian mixture model, one generally expects (1) a reduction in weight of each component, (2) a reduction in scatter, and (3) a shift of the means as the new component displaces the old. 
Figure~\ref{fig:Bz_Kcf} compares red fraction, RS mean $g-r$ color, and RS $g-r$ scatter for varying component counts $K = \{2, 3, 4\}$ in the Buzzard flock. 
We find that while the RS has a relatively consistent mean and scatter on adding components, the RS is subdivided between components at low redshifts, resulting in a drastically different weight for the reddest component. 

We find good consistency in red fraction (leftmost plot of Figure~\ref{fig:Bz_Kcf}) for $z > 0.4$, with only minor reduction in weight with added components (typically order 1\%, consistently $<5\%$). 
For $z < 0.4$, we see a severe reduction in red fraction of the reddest components for $K=3$ and $K=4$, due to the RS being subdivided into multiple components. 

\begin{figure}\centering
  \includegraphics [width=\linewidth] {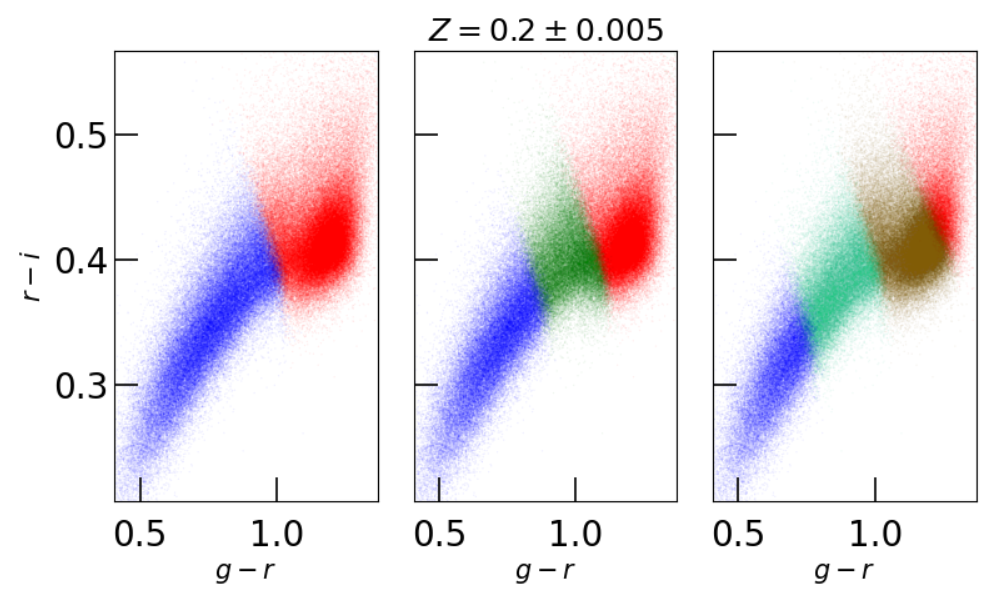}
  \caption{
    Red Dragon population characterization for Buzzard galaxies in the thin redshift slice $z=0.2 \pm 0.005$ for different component counts. {\bf Left:} $K=2$, {\bf Middle:} $K=3$, {\bf Right:} $K=4$. With four components, the RS and BC are further subdivided into bluer and redder components. 
  }
  \label{fig:Zp2_Bz_NK}
\end{figure}

At a single redshift slice ($z \approx 0.2$), Figure~\ref{fig:Zp2_Bz_NK} shows this sub-division of the RS for each component count. On the same axes of $g-r$ vs $r-i$, the thin sample of galaxies are colored by component classification for $K=2$ to $K=4$ component mixtures. Addition of components primarily splits populations into bluer and redder sub-populations, but the fraction of star-forming or quiescent galaxies belonging to each population varies strongly with changing $K$ values. 

\begin{figure}\centering
  \includegraphics [width=\linewidth] {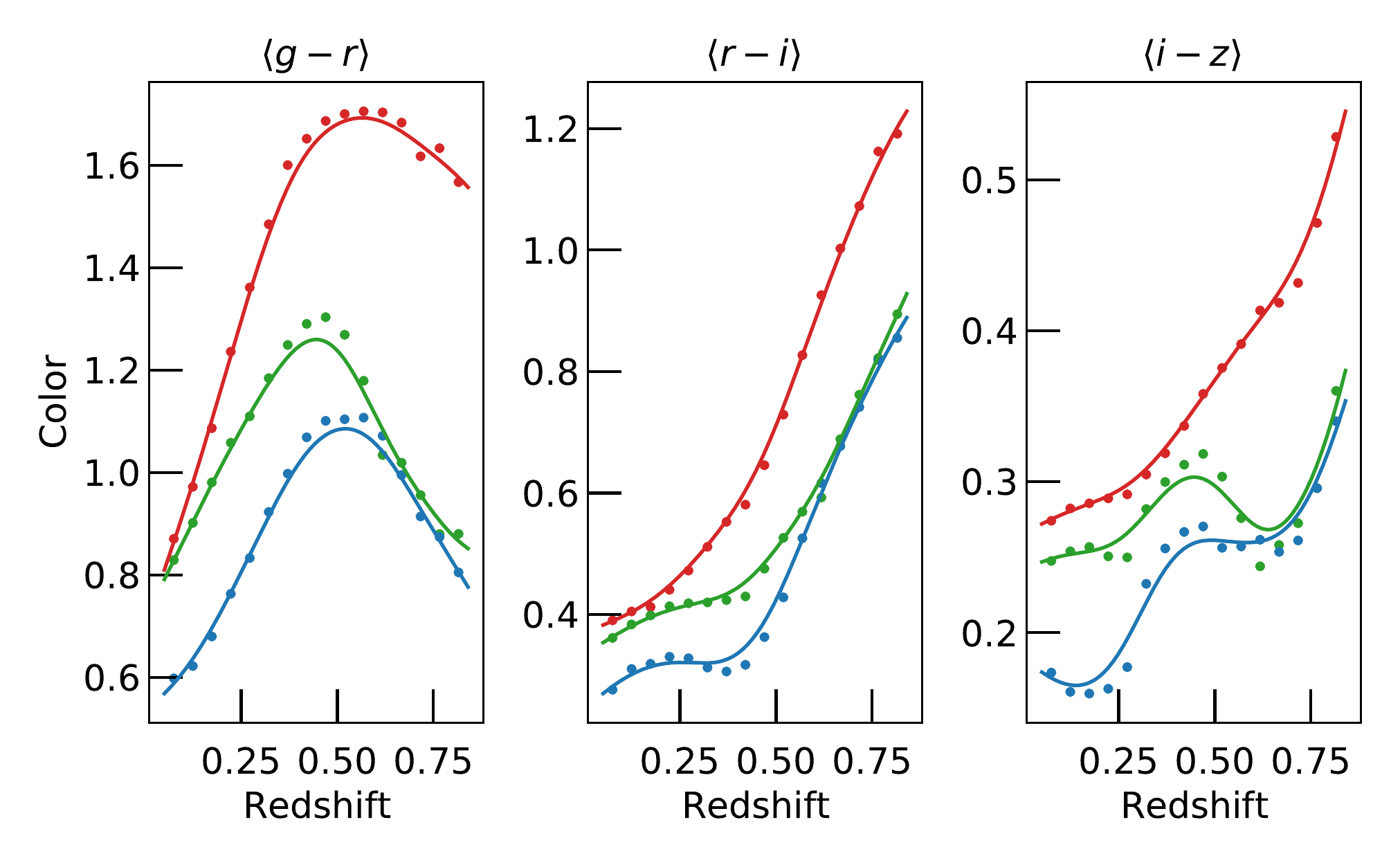}
  \caption{
    Mean color values for all three components of a $K=3$ fitting of Buzzard photometry. The red lines consistently track a quiescent population and the blue lines consistently track the star-forming population. The green component moves from clearly matching the star-forming population at high redshifts to clearly matching the quiescent population at low redshifts. 
    % This illustrates a danger of using more than two components. 
  }
  \label{fig:Bzz_3K_mu}
\end{figure}

Figure~\ref{fig:Bzz_3K_mu} illustrates this sub-division in mean colors $\mu_\alpha(Z)$ for the $K=3$ model across redshift, similar to Figure~\ref{fig:Bz_mu}: the middle component (green line) matches the BC at high redshifts, but matches the RS at low redshifts. 
Though this could be characterizing galaxy evolution from BC to RS, it may be a statistical artifact, where the RS is more non-Gaussian at low redshifts than at high redshifts, as compared to the BC. Without data on sSFR and galaxy evolution in Buzzard, it's difficult to say. 
If using $K>2$, one must be careful in labeling the RS, since the quiescent population may be split between components. 

Besides weight, we find excellent consistency in the other GM parameters. Mean color changed on average by $\lesssim 0.01~{\rm mag}$ on adding a new component (consistently less than $0.05~{\rm mag}$; see Figure~\ref{fig:Bz_Kcf}, center plot). 
Scatter reduced on average by $<5\%$ with each added component (consistently less than $<33\%$; see Figure~\ref{fig:Bz_Kcf}, rightmost plot). 
Correlations varied by less than 12\% on average. 
% In contrast to the RS, the BC is subdivided by the added components, drastically altering its parameterization and complicating comparison. 
Though the populations are consistently characterized, single components may not consistently correspond to the same population; middle components may at one redshift characterize the quiescent population but at another characterize the star-forming population. 

% [role of non-Gaussianity + summary; something like:]
As discussed earlier (see \S\ref{sec:elment_count}), the optimal count of Gaussian components to use in modeling depends on the dataset; increased non-Gaussianity requires more nuanced modeling. Despite non-Gaussianities in our datasets, the two-component model still aptly characterizes the galaxy population. Adding extra components must be done with care due to complexities in modeling and interpretation. 
\section{Conclusions} \label{sec:conclude}

We present Red Dragon, a new method for galaxy population characterization, which evolves Gaussian mixture models in the space of broad-band optical colors across redshift. With the red sequence of quiescent galaxies as a target population in both observed and simulated galaxy samples, we demonstrate the method's ability to identify the quenched population with similar accuracy to previous approaches but with smoother continuity across redshift (in addition to characterizing the BC). 

Jumping from using one color alone to another as a RS selector (e.g the transition from $g-r$ to $r-i$ near $z=0.4$) gives an inherent discontinuity in red fraction  $f_R(z)$ or in accuracy of selecting the quenched population. Since Red Dragon interpolates Gaussian mixture parameters across redshifts in multi-color space, the resulting characterization of galaxy photometry yields a continuous red sequence definition (thereby resulting in a continuous red fraction). 
% \wm{[nix/reword?]: This means that a cluster richness defined by Red Dragon would remain constant were that cluster at any redshift.} 
If metrics such as richness and red fraction are to be broadly interpretable across wide redshift spans, we must move beyond discontinuous single-color selection of the RS. 

By construction, Red Dragon also offers a new way to explore RS, GV, and BC photometric behavior through its explicit fits to population weight, mean color, scatter, and correlation between colors. This fitting allows investigations into the photometric sub-populations of galaxies, such as fraction of transitioning galaxies as a function of redshift, or the evolution of mean color and scatter of blue cloud galaxies.

%This methodology results in a better fit to the galaxy photometric population, which is inherently quite Gaussian in multi-color space. 
Fitting the population with Gaussian mixtures results in similar or superior selection of quenched galaxies as optimized color–magnitude (CM) or color–color (CC) selections (see figures~\ref{fig:cf_Ncol} and~\ref{fig:bACC_CM_CC_GM}). 
Simple CM and CC selections lack information gained from the other colors (corresponding to properties such as dust, age, and metallicity; see section~\ref{sec:beyond}), which information would help disentangle degeneracies between red sequence and blue cloud in photometric space. 
% This definition of the RS was relatively consistent with increasing component count. 
Though an optimized cut in CC space can select the quenched population with similar accuracy to that of a GMM, such a selection only works well for a limited redshift span. To preserve accuracy across larger redshift spans, interpolating Gaussian mixtures in multi-color space serves as a straightforward and natural way to extend the RS to higher redshifts. 

% \wm{[nix following \P?]} First results from Buzzard show that Red Dragon’s selection of the red sequence is relatively robust to both magnitude evolution (i.e. how low and high stellar mass galaxies differ at a fixed redshift) and component count (i.e. using two components as RS + BC vs including a GV component or even a fourth component). Individual parameters of the GMM do depend on magnitude though, especially so with red fraction, which declines near linearly with magnitude. 
% Buzzard halos with richness $\lambda_{\rm RD}$ calculated from Red Dragon had less scatter at fixed halo mass than a richness calculated from more traditional CM fitting, hinting at the possibility of it being an improved mass proxy (see section~\ref{sec:proxy}). 

We note here that in addition to extending deeper into the infrared (towards $Y$, $J$, $H$ filters), higher-energy wavelengths could also be added. Even X-ray data show differences between RS and BC \citep{Comparat+22}, so pan-chromatic analyses of populations would certainly improve characterization of the RS and BC, better distinguishing the two populations.

A continuous RS definition across wide redshift spans will be critical as future galaxy surveys push deeper and fuller into the redshift regime. 
The RS has already been detected beyond $z=2$, with evidence for a quenched population out to $z<2.5$
\citep{Kriek+08,Williams+09,Gobat+11,GM+21}. 
Future telescopes such as \emph{Euclid}\footnote{
  Planned launch date: Q1 2023; its \href{https://www.euclid-ec.org/?page_id=2490} {Y, J, H filters} span 900 to 2000~nm.
} and the \emph{Nancy Grace Roman Space Telescope}\footnote{
  Planned launch date: by May 2027; its \href{https://roman.gsfc.nasa.gov/science/WFI_technical.html}{six filters} span 480 to 2300~nm. % wide field instrument (WFI)
} will provide NIR-band observations of galaxies out to high redshifts, measuring the 4000~\AA\ break for high-$z$ galaxies. 
Modeling of the RS continuously beyond $z=1$ (and eventually $z=2$) will be critical moving forward, allowing for meaningful interpretation of measures such as red fraction or richness.

%%%%%%%%%%%%%%%%%%%%%%%%%%%%%%%%%%%%%%%%%%%%%%%%%%
\section*{Acknowledgements}
% The Acknowledgements section is not numbered. Here you can thank helpful colleagues, acknowledge funding agencies, telescopes and facilities used etc. Try to keep it short.
The authors thank the Leinweber Center for Theoretical Physics (LCTP) at University of Michigan for their generous graduate fellowship which enabled this research projects' completion. 

WKB thanks Johnny Esteves, Eric Bell, Eli Rykoff, Oleg Gnedin, and Peter Melchior for their insights and assistance in this project. WKB also thanks his wife Eden and his son Fletcher, by whom his days are vastly brightened and to whom he owes his life. 

This work was made possible by the generous open-source software of \href{https://numpy.org/doc/stable/} {\sc NumPy} \citep{numpy}, \href{https://matplotlib.org/} {\sc Matplotlib} \citep{pyplot}, \href{https://docs.h5py.org/en/stable/} {h5py} \citep{h5py}, \href{https://scikit-learn.org/stable/} {\sc Sci-kit Learn} \citep{sklearn}, and \href{https://github.com/afarahi/kllr/tree/master/kllr} {KLLR} \citep{Farahi+18,Anbajagane+20}.

%%%%%%%%%%%%%%%%%%%%%%%%%%%%%%%%%%%%%%%%%%%%%%%%%%
\section*{Data Availability}
% The inclusion of a Data Availability Statement is a requirement for articles published in MNRAS. Data Availability Statements provide a standardised format for readers to understand the availability of data underlying the research results described in the article. The statement may refer to original data generated in the course of the study or to third-party data analysed in the article. The statement should describe and provide means of access, where possible, by linking to the data or providing the required accession numbers for the relevant databases or DOIs.

The current iteration of Red Dragon is available on Bitbucket at \href{https://bitbucket.org/wkblack/red-dragon-gamma/src/master/}{wkblack/red-dragon-gamma}.  
Methods used in this analysis are also available at \href{https://bitbucket.org/wkblack/red-dragon/src/master/}{wkblack/red-dragon} on Bitbucket. 
Data from \href{http://skyserver.sdss.org/dr16/en/tools/search/sql.aspx}{SDSS} and \href{https://www.tng-project.org/data/downloads/TNG300-1/}{TNG} are publicly available on their respective servers.


\bibliographystyle{mnras}
\typeout{} % https://www.overleaf.com/learn/latex/Questions/BibTeX_isn't_working;_my_%5Ccite_are_showing_up_as_question_marks_(%3F) suggested I add this line... 
\bibliography{main} % if your bibtex file is called main.bib
% \vfill 
% \newpage

%%%%%%%%%%%%%%%%%%%%%%%%%%%%%%%%%%%%%%%%%%%%%%%%%%
%%%%%%%%%%%%%%%%% APPENDICES %%%%%%%%%%%%%%%%%%%%%

\appendix

\newpage
%%%%%%%%%%%%%%%%%%%%%%%%%%%%%%%%%%%%%%%%%%%%%%%%%%
\section{SDSS vs TNG color distribution} \label{apx:SDSS_vs_TNG_hist}

\begin{figure*}\centering
  \includegraphics[width=\linewidth]{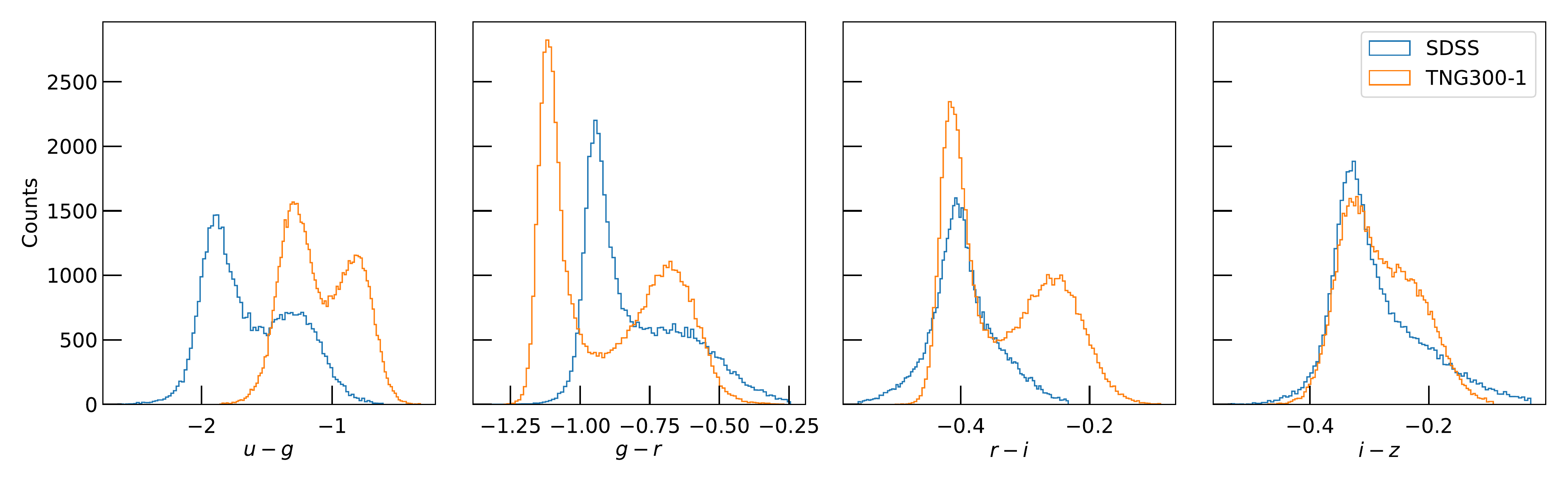}
  \caption{
    Comparative number densities in SDSS (blue) vs TNG (orange; simulated errors) for primary colors (both at $z \doteq 0.1$). Note that for each color, TNG has far more drastic valleys between RS and BC than SDSS (which only visibly shows a valley for $u-g$ here). Thus, any selection of RS and BC in TNG will be far cleaner than in SDSS. 
  }
  \label{fig:hist_SDSS_vs_TNG}
\end{figure*}

Though photometry between SDSS and TNG have similar mean colors, the distinctness of RS and BC differ significantly between the two datasets. 
Figure~\ref{fig:hist_SDSS_vs_TNG} shows histograms of color distributions for TNG as compared to SDSS. Errors on TNG magnitudes were simulated to match SDSS trends, as detailed in table~\ref{tab:SDSS_errs}. 
\begin{table}\centering
  \begin{tabular}{c|l l}
    band & log slope $\alpha_x$ 
         & intercept $\epsilon_{0,x}$ \\ 
    \hline 
    $u$  & $0.235 \pm 0.001$ & $-1.742 \pm .0019$ \\
    $g$  & $0.185 \pm 0.001$ & $-2.212 \pm .0004$ \\
    $r$  & $0.187 \pm 0.001$ & $-2.182 \pm .0004$ \\
    $i$  & $0.195 \pm 0.001$ & $-2.097 \pm .0006$ \\
    $z$  & $0.246 \pm 0.001$ & $-1.619 \pm .0011$ 
  \end{tabular}
  \caption{
    Log slope $\alpha_x$ and intercept $\epsilon_{0,x}$ of SDSS typical band errors, fitting $\log_{10} \epsilon_x = \alpha_x (m_x - m_{\rm ZP}) + \log_{10} \epsilon_{0,x}$ for each band $x$, with magnitude zero point $m_{\rm ZP} = 17.3$. 
    Note that these are fairly close to error $\epsilon \propto L^{-1/2}$, which implies $\log_{10} \epsilon = .2 \, m_x + C$. 
  }
  \label{tab:SDSS_errs}
\end{table}
Even with these added errors, TNG displays far clearer separations between RS and BC than SDSS. While SDSS only has a visible valley for $u-g$ (in the primary colors), TNG shows significant valleys for all but $i-z$ (the noisiest bands, furthest from the 4000~\AA\ break at this low redshift). 

The stronger dichotomies of TNG made selection of the RS far cleaner for TNG than for SDSS. 
This explains the vast increase in constraining power in Figure~\ref{fig:bACC_CM_CC_GM} from SDSS to TNG, and the relatively consistent accuracy across selection methods.

%%%%%%%%%%%%%%%%%%%%%%%%%%%%%%%%%%%%%%%%%%%%%%%%%%
% \section{Main astrophysical determiners of galaxy spectra}
\section{Galaxy Spectra Astrophysics}
\label{apx:astrophysics}

\begin{figure}\centering
  \includegraphics [width=\linewidth] {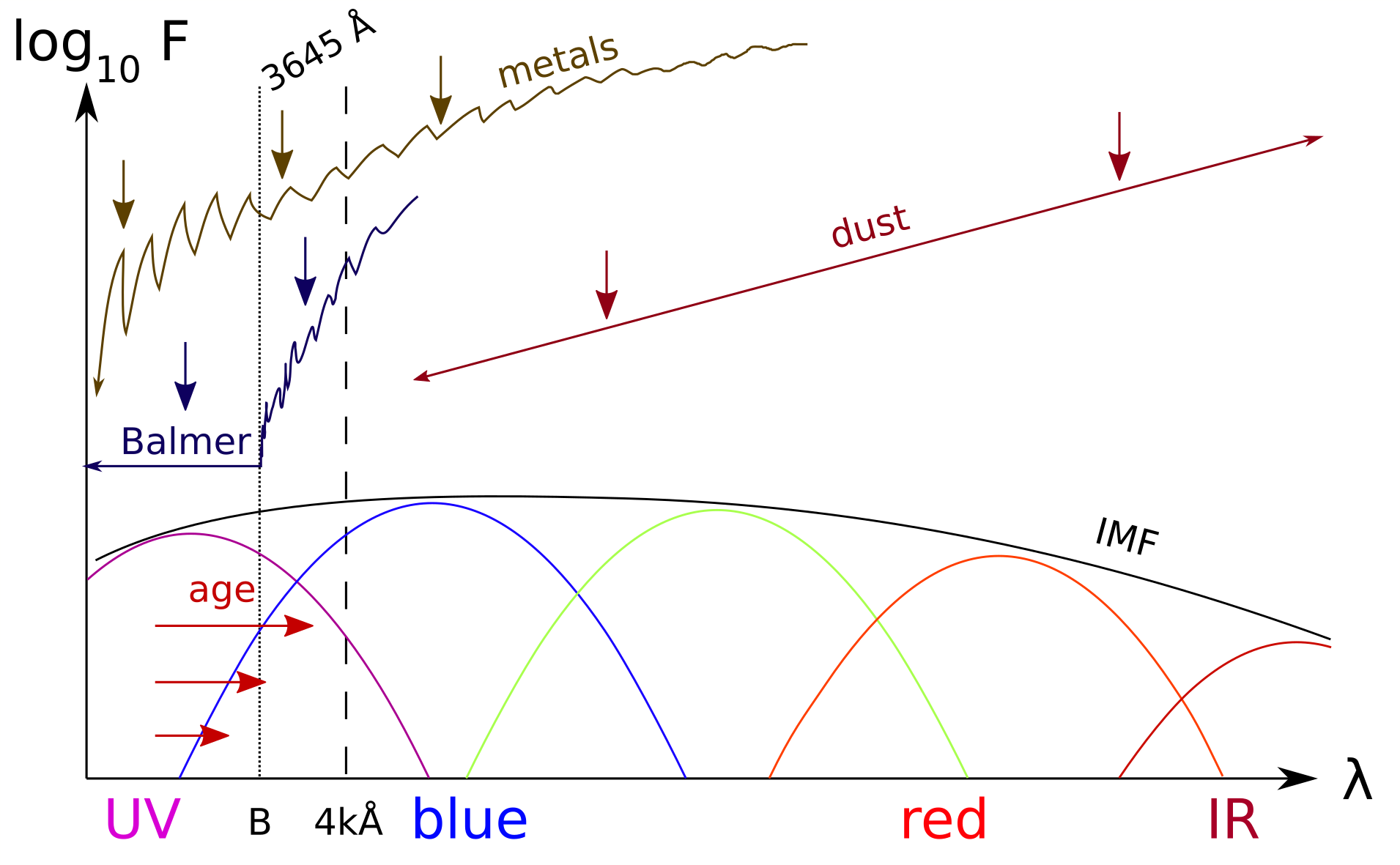}
    \caption{
      Summary cartoon of main astrophysical effects determining galaxy colors. 
      % % % % % % % % % % % % 
      % emission: 
      Stars provide the vast majority of optical light, with short-lived blue giants providing the majority of short-wavelength light and longer-lived red giants providing the majority of longer optical wavelengths. 
      The overall shape of the emission spectrum depends on both the stellar initial mass function (IMF) and the star formation history (SFH). 
      % age
      Main sequence stars redden with age, with blue giants aging the fastest ($\tau_{\rm MS} \propto M^{-2.5}$). Thus, old galaxies emit less short-wavelength (optical) light. 
      % % % % % % % % % % % % 
      % absorption: 
      Primary sources of absorption include dust reddening and line absorption. 
      % dust 
      Dust presence causes Rayleigh scattering: the preferential scattering of short-wavelength light (intensity $I \propto \lambda^{-4}$). Such scattering reddens the overall spectrum, which can be clearest observed in the post-break slope. 
      Though this effect is somewhat degenerate with age and metallicity, both age and metallicity reduce low frequencies exponentially more, whereas dust effects are linear in log flux. 
      % metals
      The many absorption lines of various metals convene around (and especially at wavelengths shorter than) 4000~\AA, resulting in a blanket of absorption. Galaxies with identical $\tau Z^{3/2}$ (i.e. $\Delta \, {\rm age} \, / \, \Delta \, {\rm metallicity} = 3/2$) have virtually identical optical colors \citep{Worthey_1994}, so the two effects are highly degenerate. 
      % hydrogen line absorption
      In contrast to the smoother effects of age or metallicity, hydrogen line absorption asymptotes towards 3645~\AA\ from above (with shorter wavelengths both fully ionizing hydrogen atoms and imparting a surplus kinetic energy kick). This convergence of the Balmer series results in a sudden drop in emissions---sharper than that of line blanketing. 
      % Summary
      Thus the IMF / SFH, age, dustiness, and metallicity of galaxies all play a role in determining their overall spectrum and membership of the RS / BC. 
    }
    \label{fig:cartoon_gal_spec}
\end{figure}

This section summarizes the main astrophysical effects which determine spectral shape for galaxies at optical wavelengths. 
Generally, the most striking spectral difference between RS and BC is the galaxy's quenched status, as measured by the strength of D4000. However, other factors help separate RS from BC, such as galactic dust content, metallicity, or age. These secondary factors affect colors beyond just the color used to measure D4000, so multi-color analysis serves to better distinguish the RS from the BC than a raw measurement of D4000. 
% High correlations between photometric colors (from 40 to 70\% correlated in the low-$z$ SDSS sample) hint that their unique information could help separate RS from BC. 

Dust plays a substantial role in altering a galaxy's position in multi-dimensional photometric color space. 
% dust 
Due to Rayleigh scattering, shorter wavelengths scatter easier, with scattering intensity $I \propto \lambda^{-4}$. 
Thus, the presence of dust depresses bluer wavelengths, inflating the overall spectrum slope $d\log F/d\lambda$, reddening all photometric colors. 
Therefore, the post-break slope (blue--infrared slope in rest frame) correlates highly with dust content. 
Furthermore, once galaxies are stripped of their dust, insufficient material exists for stellar creation, so star formation rates decline. 

Age and metallicity also redden galaxy spectra. 
% age
As stars age, they leave the stellar main sequence and become red giants. When galaxies as a whole age, they therefore also redden. Age then affects the spectrum similarly to dust, reddening overall but particularly so near the break \citep{Worthey_1994}. 
% metallicity
Stars produce metals with age, which cause increased amounts of line blanketing. This blanketing causes widespread reddening, particularly near 4000~\AA\ and shorter (ibid.). 
As mentioned earlier, since the Balmer break (asymptotically approaching 3645~\AA) is caused by hydrogen abundance while line blanketing (approximately around 4000~\AA) is caused by stellar metallicity, the two effects result in different spectrum curvatures about the break. A ratio of the two effects relates to the metallic abundance ratio [Fe/H], so curvature correlates with metallicity \citep[see also][Fig. 2]{Chalonge_Divan_1977}. 
Though spectrum curvature cannot be encapsulated in a single color, paired colors can capture spectrum curvature to some extent. 

Figure~\ref{fig:cartoon_gal_spec} shows a graphical summary of these main astrophysical effects which determine optical--IR galaxy spectral shape. Of primary relevance are the effects of the Balmer break (at 3645~\AA), dust (reddening the entire spectrum), age (most noticed as a reduction at the blue end of the spectrum), and metallicity \citep[causing line blanketing, especially $\lesssim$ 4500~\AA; see][]{Lejeune+97}.

\subsection*{Comparison to BCD stellar classification scheme}

Incidentally, the main effects affecting galaxy spectral shape (star formation, dust presence, age, metallicity) correspond well with the BCD stellar classification scheme \citep{Barbier_Chalonge_1941}. The BCD system gives a more precise (3D, continuous\footnote{
  In practice, the BCD system is closer to 2.5 dimensions, like a bent sheet of paper. The MK system can then be mapped with fair accuracy onto this bent sheet. 
  % This is perhaps why BCD never caught on. 
}) stellar classification model than the common MK system (2D, discrete). Its three distinguishing parameters measure 
\begin{enumerate}
  \item $D$, the ratio of spectrum intensities spanning the 4000~\AA\ break;
  \item $\phi_b$, the blue--violet post-break spectrum slope; and 
  \item $\lambda_1$, the midpoint location along the drop. 
\end{enumerate}
Larger values of $\lambda_1$ indicate a smoother break, implying more line blanketing than Balmer line absorption, thus correlating well with metallicity; more directly, the difference between the NIR slope $\phi_n$ (pre-break, ca. 3500~\AA) and blue--violet slope $\phi_b$ (ca. 4000--4600~\AA) shows strong correlation with metallicity \citep [see] [Fig.~2] {Chalonge_Divan_1977}. 
% see also https://www.laserstars.org/spectra/BalmerJump.html (Y.P. Varshni and J. Talbot 2006)
As galaxy luminosities arises from stellar luminosities, it should be no surprise that similar distinguishing markers should be sought after for both stars and galaxies.

\iffalse 
%%%%%%%%%%%%%%%%%%%%%%%%%%%%%%%%%%%%
\section{Qualitative magnitude running} \label{apx:qual_mag_run}
Figure~\ref{fig:cf__CC_CM} shows galaxy photometry in a high redshift slice of the Buzzard Flock simulation set, both in color--color (CC) and color--magnitude (CM) spaces.
Thin slices in magnitude space were analyzed separately to freely evolve RS and BC with magnitude. The two groups are then colored by summer and winter colorschemes respectively, with the more orange (yellow) and blue (teal) points corresponding to brighter (fainter) galaxies. This figure then shows how RS and BC centers and spreads vary with both magnitude and color, and thereby illustrating the deficiencies of 2D CC / CM cuts. 

\begin{figure}\centering
  \includegraphics[width=\linewidth] {figs/cf__CC_CM.png}
  \caption{
    Qualitative view of split in galaxy population in color--color (left) vs color--magnitude (right) spaces at $z \doteq 0.6$ for the Buzzard Flock simulation set. 
    Components isolated using two-component Gaussian mixtures segmented across magnitude bins to highlight the running of Gaussian properties with magnitude. 
  }
  \label{fig:cf__CC_CM}
\end{figure}

% In CM space, it becomes apparent that red fraction (and other mixture parameters) varies significantly with magnitude (and redshift). Thus, selection in both CC and CM spaces suffer from lack of information. 
\fi

%%%%%%%%%%%%%%%%%%%%%%%%%%%%%%%%%%%%%%%%%%%%%%%%%%
\section{Optimal Number of Components} \label{sec:K_optimal}

In this section, we consider the optimal number of components to characterize the SDSS/low-$z$ galaxy population, as determined by the Bayesian Information Criterion \citep[BIC;][]{Schwarz_1978}:
\begin{equation}
  \mathrm{BIC} \equiv k\ln(n) - 2\ln(\widehat L)
\end{equation}
BIC penalizes large parameter counts $k$ (with additional weight for increased number of data points $n$) while rewarding increased maximum likelihood $\widehat L$, such that lower BIC values indicate a superior model.
The relative likelihood of two models is proportional to $\exp(- \Delta {\rm BIC}/2)$, so values of $\Delta {\rm BIC} \gtrsim 10$ indicate significant evidence for model superiority. 
% in particular, 
% 1 sigma ~ -2.3
% 2 sigma ~ -6.2
% 3 sigma ~ -11.8
Since only relative BIC matters, we measure here $\Delta{\rm BIC}_K \equiv {\rm BIC}_K - \min({\rm BIC}_K)$: the value of BIC (for any component count $K$) relative to the minimum BIC (across all $K$). 

\begin{figure}\centering 
  \includegraphics [width=\linewidth] {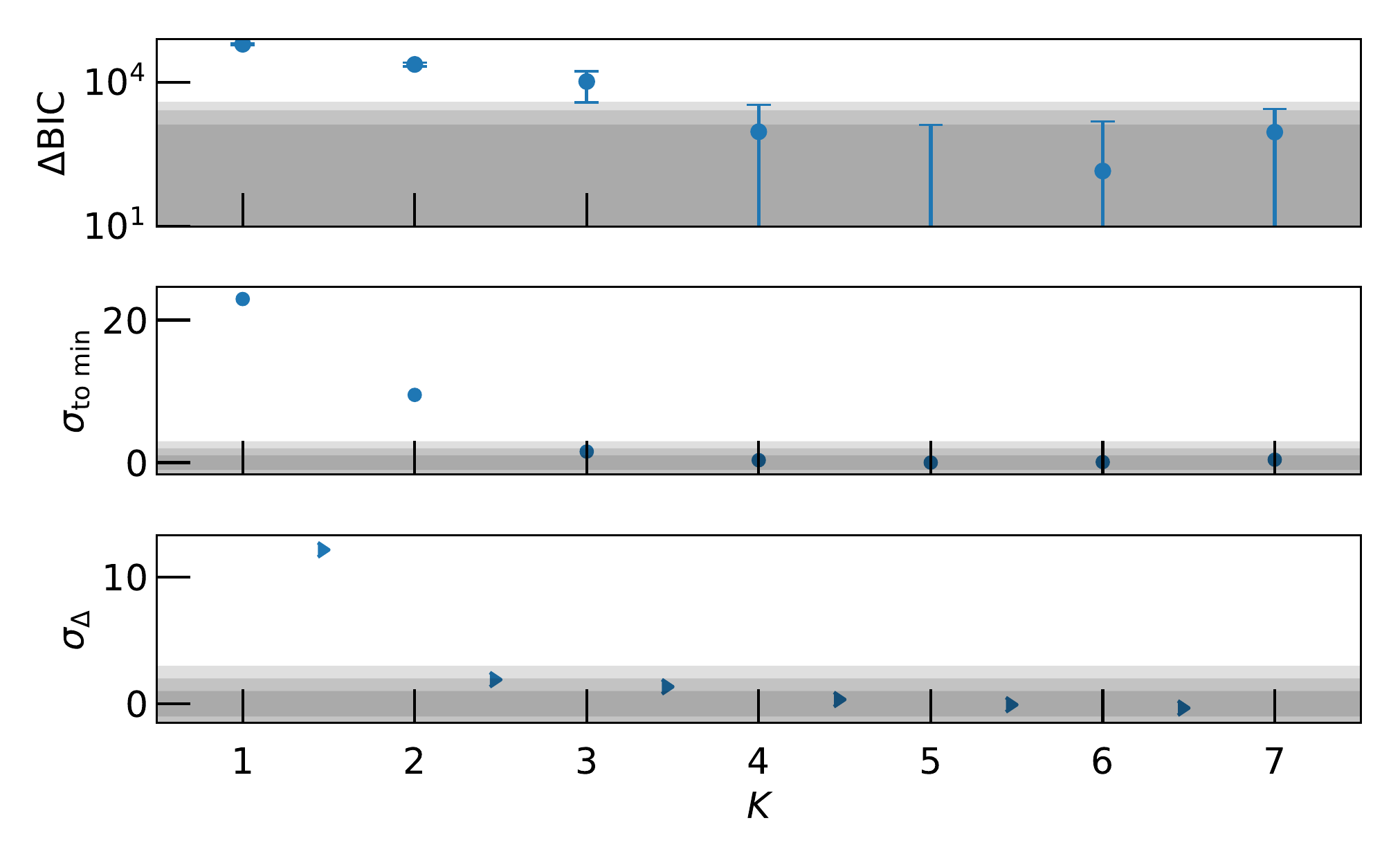}
  \caption{
    Bayesian Information Criterion (BIC) information for SDSS low redshift sample ($z=0.1 \pm 0.005$), evidencing the limited gains of $K > 3$ Gaussian mixture components. 
    Error bars estimated from bootstrap resampling. 
    Grey bands indicate $\pm 3 \sigma$ range for minimum BIC. All off-screen errorbar tails are negative, i.e. below the minimum BIC's mean value). 
    {\bf Top:}
      BIC relative differences (i.e. ${\rm BIC} - \min({\rm BIC})$). Though a decrease in BIC $>10$ indicates significant evidence for an improved model, bootstrap errors on BIC measurements reveal insignificant gains, especially at higher component counts. 
    {\bf Mid:}
      % Significance from minimum: [11.03  4.77  3.67  1.91  1.07  0.76  0.    0.18  0.31  0.63  0.91  1.34] (sklearn) 
      % [22.95  9.5   1.55  0.34  0.    0.07  0.41] (pygmmis) 
      Sigma from each point to the BIC minimum.
    {\bf Bot:}
      % Significance from adjacent: [ 3.38  4.05  1.65  0.78  0.45  0.76 -0.18 -0.14 -0.38 -0.26 -0.37] (sklearn) 
      % [12.13  1.9   1.34  0.34 -0.07 -0.34] (pygmmis) 
      Significance of BIC gain on incrementing component count. 
      % Only moving from $1 \rightarrow 2$ components decreased BIC significantly.
  }
  \label{fig:BIC_comparative}
\end{figure}

Figure~\ref{fig:BIC_comparative} shows our analysis of BIC for the SDSS/low-$z$ sample. The upper panel gives BIC values relative to the minimum BIC (at $K=5$), the middle panel shows the significance of the differences in those BIC values, and the lower panel shows the incremental gain in significance on increasing $K$. In each panel, the error of the $K=5$ model is shown for $\pm 3 \sigma$ as transparencies. These plots aid in choosing a number of components $K$ that both optimizes fitting of the galaxy population while avoiding overfitting. 

Though the upper panel shows that the SDSS/low-$z$ three-component model ($K = 3$) has $\Delta{\rm BIC} \doteq 10^{4.01}$, the middle panel shows that accounting for uncertainty at both $K=3$ and at $K=5$ (the absolute minimum) implies that the two are $< 2 \sigma$ different---the BIC of the $K=3$ model is indistinguishable from the measured minimum, statistically speaking. From a BIC standpoint, there is no statistically significant reason to use $K \geq 4$ for this dataset. 
This highlights the importance of calculating error bars on BIC values; while such a high $\Delta$BIC would indicate overwhelming support of a five-component model over a three-component model, the errors on each BIC value reveal the actual insignificance of their relative $\Delta{\rm BIC}$. 

One can also compare BIC gains incrementally: as we add an additional component, how significantly do we decrease BIC? If adding a component doesn't significantly decrease BIC, it may not be worth the added complexity. 
Though moving from one component to two gives a $> 12 \sigma$ improvement in BIC, moving from two to three components only improves BIC by $< 2 \sigma$ (with further increments yielding even smaller gains). Even though the $K=2$ model has a BIC significantly ($\sim 10\sigma$) above the absolute BIC minimum, incremental gains beyond $K=2$ are insignificant. 
This again favors models with $K < 5$, despite $K = 5$ being the BIC minimum on average. 
Based on the above analysis for this particular dataset, we advise using either $K=2$ or $K=3$ components to model the distribution of galaxies. 

Though BIC accounts for the cost of extra parameters, it doesn't account for human usefulness directly; models with fewer components are more easily analyzed whereas larger component counts become increasingly difficult to consistently model and interpret across redshift.

The spirit of BIC is to maximize likelihood while minimizing complexity: a model should use the lowest component count possible while still achieving significant decreases in BIC. 
To this end, we suggest using a two or three-component model of galaxy colors. 
Using a two-component model has the upside of easier interpretation. While two Gaussian components unambiguously model the RS and BC across all redshifts, even a three-component model can give rise to significant ambiguity. As discussed in section~\ref{sec:RS_by_K}, while the third component sometimes models a middle population (galaxies transitioning from BC to RS) at other times it models the larger scatter surrounding the RS and BC (modeling kurtosis of the population). 
Extra components beyond three lack intuitive explanation from the underlying astrophysics and lack statistical power to increase model favorability. 
Practically speaking, larger numbers of evolving components are far trickier to map between redshifts. Continuous modeling of four components is taxing at best, with meaningful interpretation of each component across all redshifts a tricky issue. {\it Caveat emptor.} % Draco Dormiens Nunquam Titillandus 
For simplicity of discussion and comparison, we employ a two-component model for our results section as a minimum-use case.

% Optimal fit component count depends on input dataset, so we give 

%%%%%%%%%%%%%%%%%%%%%%%%%%%%%%%%%%%%%%%%%%%%%%%%%%
\section{Magnitude trends in Buzzard} \label{apx:quan_mag_run}

Here we investigate trends of magnitude running as found in Buzzard. Though running of parameters $\theta$ with magnitude is statistically significant (section~\ref{sec:lintrend}), slope of mean color relative to the scatter of the populations was relatively small (section~\ref{sec:mean_shift}), so differences in selection were relatively minimal (section~\ref{sec:sim_sel}). 
Magnitude trends had the largest effect for dimmer galaxies, resulting in $\mathcal{O}(10\%)$ differences in selection.

% % % % % % % % % % % % % % % % % % % % % 
\subsection{Strength of linear trend} \label{sec:lintrend}

To give a general feel on the significance of magnitude running in Buzzard, we let each GMM parameter run linearly with magnitude and analyzed several redshift slices. 
First results show that brighter galaxies are more likely red ($d f_R / d {m_i} \approx -0.13$, such that faint galaxies are almost always BC members), with the degree of redness increasing gradually with luminosity ($d \mu_{(g-r), \, {\rm RS}} / d{m_i} \sim -0.04$), more significantly so for the BC ($d \mu_{(g-r), \, {\rm BC}} / d{m_i} \sim -0.1$). 
The RS scatter increases at fainter magnitudes ($d \sigma_{(g-r), \, {\rm RS}} / d{m_i} \sim z/3$), especially at higher redshifts, but the scatter of BC galaxies doesn't show significant linear evolution with magnitude. 
While $f_R$ runs near linearly with magnitude at a fixed redshift, $\mu$ and $\log \sigma$ have more quadratic or cubic running with magnitude (at fixed redshifts)---their fitting deserves more attention in the future, to best quantify the running of Gaussian mixture parameters with magnitude and redshift.

% # fR varies far more with magnitude than with redshift cf the other fields, followed by 'gmr_BC', then 'sig_BC' = 'sig_RS', with 'gmr_RS' varying the least with magnitude, compared to redshift.

% significantly non-zero slopes: fR > sig_RS > gmr_BC > gmr_RS > sig_BC

Red fraction varied more significantly with magnitude compared to the other parameters, with a roughly linear evolution with magnitude and redshift:
\begin{equation} \label{eqn:fR_mag_run}
  f_R(m_i,z) \sim -.13 (z+.5) (m_i - 23.1). 
\end{equation}
This shows that fits gave null $f_R$ for $m_i > 23.1$, i.e. only bright red galaxies truly belong to the red sequence---dim red galaxies belong to the blue cloud. Over Buzzard's redshift range of $z|[.05,.75]$, this means the red sequence went from 0 to 100\% across some 6--14 magnitudes, or a factor of $10^{2.5}$--$10^{5.6}$ in luminosity. 
% SDSS 3.5--4 m_i range
Thus a linear fit to the RS catches the most BC members at faint magnitudes. 

Allowing $f_R$ to run with magnitude could produce a more natural definition of the RS than traditional luminosity-limiting definitions of richness $\lambda$. 
Since faint galaxies rarely belong to the RS (while the brightest galaxies nearly always belong to the RS), a negative value of $d f_R / d {m_i}$ gives an inherent cutoff to the RS (as compared to the contrived cutoff of e.g. $L > 0.2 \, L_*$ or $m < m_* + 1.75$). Defining RS galaxies in such a way would correspond better to the underlying photometric distribution than defining the RS with hard magnitude cuts, potentially improving its power as a mass proxy and its universality across redshift.

% % % % % % % % % % % % % % % % % % % % % 
\subsection{Shift of mean color relative to RS width} \label{sec:mean_shift}

Though the RS may have a statistically significant shift in mean color, it may not significantly change selection of the RS by Gaussian mixture if that running is small compared to the width of the RS. 
To measure this, we use the metric $\varsigma$ to quantify slope of the RS relative to its scatter: 
\begin{equation}
  \varsigma \equiv \frac{d \langle \mu_a \rangle / dM_b} { \sigma_a } % \wm{\: \text{if varsigma not okay, then I'll use tau.}}
\end{equation}
(for color $c_a$ and magnitude $M_b$; the magnitude may be any---our three fits use $M_r$, $M_i$, and $M_z$). 
%\begin{equation}
%  \varsigma \equiv \frac{d \langle \mu_j \rangle / d M_x} { \sigma_j } 
%\end{equation}
%(where the mean $\mu$ and scatter $\sigma$ correspond to the same color; the magnitude may be any band, though here our fits use $M_r$, $M_i$, and $M_z$). 

For a sample of $0.2 \, L_*$ limited galaxies with luminosities following a Schechter function (with $\alpha|[-1.5,-1]$), $3\sigma$ worth ($>99.7\%$) of its galaxies fall in a magnitude spread of $\Delta{\rm mag} < 3.3$ and $5\sigma$ worth ($>99.9999\%$) of its galaxies fall in in $\Delta{\rm mag} < 4.4$. Thus the shift in mean color generally moves less than four magnitudes. 
Using the $\pm 2 \sigma_{\rm RS}$ definition of RS width from \citet{Hao+09}, we can then take $4 \sigma$ as the distance needed to move such that the RS radically shifts with magnitude. 
These two factors of four essentially cancel out, leaving $\varsigma$ as a unitless metric for significance of RS running. If $|\varsigma| < 1$, the magnitude variation of the RS mean color is completely within the typical scatter of the RS, but if $|\varsigma| > 1$, then the RS mean color moves beyond the typical scatter of the RS. 

We measured $\varsigma$ values from SDSS data \citet{Baldry+04}, the Buzzard flock, and a redMaPPer \citet{Rykoff+14} fit of the DES~Y3 RS. 
Each of the three datasets had typical values of $|\varsigma_{\rm RS}| \lesssim 0.5$ and for all cleanly measured cases (low-magnitude fits on Buzzard were questionable) had $|\varsigma_{\rm RS}| < 1.3$, implying that in nearly every case, the running of the RS mean color was small compared to its width. 
This implies that it takes many magnitudes to significantly shift the mean RS color, relative to its intrinsic scatter. Therefore, characterizing the RS in a magnitude-ignorant way will still properly select galaxies. 

% From the fits of \citet{Baldry+04}, $\varsigma$ (using $M_r$) is usually $-0.5 \pm 0.3$ but ranges down to about $-1$, indicating that the shift of mean color over a magnitude is generally half of the $1\sigma$ scatter of the population. It takes approximately six magnitudes for a $3\sigma$ shift in RS mean color. 

% Our fits on Buzzard data (using $m_i$) show similar trends, with $\varsigma_{\rm BC} \sim -0.5$ and $\varsigma_{\rm RS} \sim 0$ until fainter magnitudes (where $\varsigma$ may become more negative, though the trend was inconsistent across various colors or redshifts). 

% % % % % % % % % % % % % % % % % % % % % 
\subsection{Similarity in selection} \label{sec:sim_sel} 

Though including magnitude running of GMM parameters would better represent the photometric population, the extra dimensionality of running with magnitude makes fitting more of a nightmare, and doesn't drastically alter selection. 

Though the current version of Red Dragon doesn't include magnitude running, there are a few workarounds, if you're hell-bent on including its effect. 1) You can slice your data by magnitude and run Red Dragon to characterize a bright vs dim sample of your galaxies, then interpolate between (or simply bin using) the two fits to estimate $P_{\rm red}$ for individual galaxies. 
2) If working with a relatively thin redshift slice, you can set in the code \verb`Z=m_i`, i.e. let Red Dragon measure and interpolate across magnitude rather than redshift. 
We use this latter method to measure similarity in selection between magnitude-running and redshift-running versions of Red Dragon. 

At several thin redshift slices (for $z[.1,.7]$), we measured $P_{\rm red}$ differences between the color-only and magnitude-running dragons. 
In binary selection, these two dragons resulted in about 5\% opposing characterization (i.e. 5\% of galaxies were characterized as RS instead of BC or vice versa) with a balanced accuracy of about 95\% (between magnitude-running dragons and redshift-running dragons; not an accuracy of selecting the quenched population). 
Looking at $\Delta P = (P_{\rm red,mag} - P_{{\rm red},z})$, we found that though about $\sim 7\%$ of galaxies had $|\Delta P| \gtrsim 0.25$, less than about 1.5\% of galaxies had $|\Delta P| \gtrsim 0.50$ (so few galaxies had significantly different characterization). There was approximately 12\% scatter off $\Delta P = 0$. 
So while the effect of running with magnitude is notable, it can be ignored without much loss. 
% Those numbers were consistent to less than a percent using linear vs spline interpolation (for Z=.3 old version). 

Magnitude matters most when you're uncertain of redshift, but the current version of Red Dragon is designed for clusters, where you have very good redshift estimates. 
For the current iteration of Red Dragon, the algorithm remains magnitude-ignorant, solely using colors to distinguish RS from BC.

% % % % % % % % % % % % % % % % % % % % % 
\section{Redshift selection} \label{sec:Z_selection} 

To avoid overwhelming pyGMMis, galaxies are chosen probabilistically (with replacement, allowing for bootstrap analyses). If redshift errors are supplied, then probability of selecting a given galaxy is weighted by its chance of being in a given redshift bin. 
Assuming Gaussian errors, the probability of some value $\mu \pm \sigma$ lying in interval $(a,b)$ is
\begin{equation}
  P[(\mu \pm \sigma) \in (a,b)] = \frac12 \left| \erf \left( \frac{b-\mu} {\sqrt{2} \, \sigma} \right) - \erf \left( \frac{a-\mu} {\sqrt{2} \, \sigma} \right) \right|.
\end{equation}
So we select galaxies randomly from the full sample, weighted by that probability. 

If redshift errors are not available, then we randomly pick $N$ galaxies (with replacement, allowing for bootstrapping) within the desired redshift interval.

\section{Excluding data outliers} \label{sec:outlier_exclusion}
We provide users and option to cut out extraneous galaxies based on a cut of how many `sigma' anomalous a galaxy is. This is useful for noisy datasets with notable counts of extreme outliers, be it due to either improper redshift estimates or erroneous photometry. 

First measure for each color $a$ the mean $\mu_a$ and scatter $\sigma_a$ of the population in a given redshift selection. Then for each galaxy $i$ (with color error $\delta_{i,a}$), measure how many `sigma' anomalous it is: 
\begin{equation}
  N_{\sigma,i,a} = \frac{c_a-\mu_a}{\sigma_a + \delta_{i,a}}
\end{equation}
Then sum across all colors to get a total anomaly: 
\begin{equation}
  N_{\sigma,i,{\rm tot}}^2 = \frac{1}{N_{\rm col}} \sum_a N_{\sigma,i,a}^2
\end{equation}
This unitless value then catches outliers be they of any color. If multiple colors have a high $N_{\sigma,a}$ then the value of $N_{\sigma,{\rm tot}}$ will be far larger accordingly. 
This usually removes far less than 1\% of the population, though it varies somewhat from dataset to dataset and from redshift to redshift. 
% old definition tossed <3\%, under new definition tosses <<1\%. 

%%%%%%%%%%%%%%%%%%%%%%%%%%%%%%%%%%%%
\section{Component continuity} \label{apx:continuity}
In order to aid in matching components across redshift bins, we present here scalar metrics $C_\alpha$ and $S_\alpha$ which normalize and summarize information from the mean colors $\vec \mu_\alpha$ and the scatters $\vec \sigma_\alpha$ for each component $\alpha$. These reduce the dimensionality of the problem of matching components across redshift. 

First, we estimate total population variance in each color at a given redshift slice. Since the parameter fits lack this information from the native dataset, this is estimated from a weighted sum across fitted components. 
\begin{equation}
  {\vec{\sigma}^2}_{\rm tot} \equiv \sum_\alpha w_\alpha \, \vec{\sigma}^2_\alpha
\end{equation}
We can then estimate the mean fraction of the total scatter at that redshift each component makes up:
\begin{equation}
  S_\alpha \equiv \left\langle \sqrt{ \vec{\sigma}^2_{\alpha} \big / \vec{\sigma}^2_{{\rm tot}}} \right\rangle 
\end{equation}
(averaging across each color). This then gives us a scalar estimator for the relative scatter of each component.

Second, we estimate total population mean color, similarly using a weighted combinations of the various components. \begin{equation}
  \vec{\mu}_{\rm tot} \equiv \frac{1}{N_{\rm col}} \sum_\alpha w_\alpha \, \vec{\mu}_\alpha 
\end{equation}
We then re-scale mean colors of each component relative to $\mu_{\rm tot}$ as a fraction of $\sigma_{\rm tot}$: \begin{equation}
  C_\alpha \equiv \frac{1}{N_{\rm col}} \sum_a \frac{ \vec{\mu}_{\alpha,a} - \vec{\mu}_{\rm tot} }{ \vec{\sigma}_{{\rm tot},a} }
\end{equation}
(again averaging across all colors). 

These scalar summary variables $C_{\alpha,i}$ and $S_{\alpha,i}$ then give the mean relative color and scatter of each component $\alpha$ at a given redshift slice $i$. 
The RS will generally be the reddest and least scattered component in any color, so larger $C-S$ values generally belong to the RS component whereas smaller $C-S$ values generally belong to the BC. 
Though this method is relatively straightforward for two-component models, the relationship between $C$ and $S$ for distinguishing components becomes more of an art form with higher component counts. Altering the relative importance of scatter and color can better distinguish components in some circumstances. 
(For example, $C + \frac12 S$ disentangled components better for $K=3$ Buzzard data than $C-S$.) 
Each dataset and each parameterization may behave differently, so we allow a malleable sorting metric $x = C + f \cdot S$ with user-chosen factor $f$ to determine component matching. 

If such automation of sorting is unfeasible, then the user always has at last resort sorting by hand, manually determining redshift continuity of components. This metric of $x \equiv C + f S$ is only meant to help automate the lion's share of an intricate and many-dimensional problem into a more manageable one.

\subsection{Excluding odd redshifts from fit}

In some rare occasions, parameterizations may look extremely dissimilar from their neighboring fits. 
Such a situation may arise due to a cluster of photometric outliers, odd redshift bins, or data which could lead to various Gaussian mixture parameterizations (where the GMM chose to fit dissimilarly from neighboring bins). 
These oddities may be excluded from the overall interpolation; by default, if colors are $5\sigma$ anomalous from their neighbors, they're excluded from the interpolation. 

Outliers are defined by looking at the (finite difference) curvature in colors from bin to bin. (That is, if a point lies on the line connecting its neighbors, it has zero curvature, but if it's far off the line, it's an outlier with strong curvature.) If, relative to all other redshift bins, one has $>5\sigma$ anomalous curvature, it is excluded by default (though it may be re-admitted by hand).

\vfill 
%%%%%%%%%%%%%%%%%%%%%%%%%%%%%%%%%%%%%%%%%%%%%%%%%%
% Don't change these closing lines: 
\bsp % typesetting comment
\label{lastpage}
\end{document}